\documentclass[11pt]{article}
\usepackage{amsmath}
\usepackage{amssymb}

\makeatletter

\usepackage{latexsym}
\usepackage{amsfonts}

\@addtoreset{equation}{section}

\newcommand{\eref}[1]{(\ref{#1})}

\def\be{\begin{equation}}
\def\ee{\end{equation}}
\def\ba{\begin{eqnarray}}
\def\ea{\end{eqnarray}}
\def\bet{\begin{tabular}}
\def\eet{\end{tabular}}
\def\pa{\partial}

\def\nn{\nonumber}
\def\ve{\varepsilon}

\def\a{\alpha}
\def\bt{\beta}
\def\g{\gamma}
\def\G{\Gamma}
\def\D{\Delta}

\def\dl{\delta}
\def\la{\lambda}

\def\s{\sigma}
\def\m{\mu}
\def\n{\nu}
\def\ra{\rightarrow}
\def\vp{\varphi}

\long\def\symbolfootnote[#1]#2{\begingroup%
\def\thefootnote{\fnsymbol{footnote}}\footnote[#1]{#2}\endgroup}

\makeatother

\begin{document}
\begin{titlepage}

\global\long\def\LimV{\underset{V\rightarrow1}{\mathrm{Lim}}\;}
\global\long\def\LimE{\underset{\varepsilon\rightarrow0}{\mathrm{Lim}}\;}

\begin{flushright}
Preprint DFPD/2014/TH09\\
July 2014\\

\par\end{flushright}

\vspace{0truecm}

\begin{center}
\textbf{\Large Electrodynamics of massless charged particles \vskip0.3truecm}
\par\end{center}{\Large \par}

\begin{center}
\vspace{0.2cm}

\par\end{center}

\begin{center}
Kurt Lechner\symbolfootnote[2]{kurt.lechner@pd.infn.it}
\par\end{center}

\vskip1truecm

\begin{center}
\textit{Dipartimento di Fisica e Astronomia, Universit\`a degli
Studi di Padova, Italy}
\par\end{center}

\begin{center}
\textit{and}
\par\end{center}

\begin{center}
\textit{
 INFN, Sezione di Padova,}
\par\end{center}

\begin{center}
\textit{Via F. Marzolo, 8, 35131 Padova, Italy}
\par\end{center}

\vspace{0.3cm}

\begin{abstract}
\vskip0.2truecm

We derive the classical dynamics of massless charged particles in a rigorous way from first principles. Since due to ultraviolet divergences this dynamics does not follow from an action principle, we rely on a) Maxwell's equations, b) Lorentz- and reparameterization-invariance, c) local conservation of energy and momentum.
Despite the presence of pronounced singularities of the electromagnetic field along  Dirac-like strings, we give a constructive proof of the existence of a unique distribution-valued  energy-momentum tensor. Its conservation requires the particles to obey standard Lorentz equations and they experience, hence, no radiation reaction. Correspondingly the dynamics of interacting classical massless charged particles can be consistently defined, although they do not emit {\it bremsstrahlung} end experience no self-interaction.

\vspace{0.1cm}

\end{abstract}
\end{titlepage}

\newpage

\baselineskip 5mm

\section{Introduction}

The existence of massless charged particles in nature is still an open problem, from a theoretical as well as - in a certain sense - experimental point of view. We may indeed look on gluons and gravitons as ``charged'' particles, respectively under strong and gravitational forces, and the latter are actually supposed to exist as free particles, not subject to confinement.

From a theoretical point of view, and especially in the context of electromagnetic interactions, in quantum theory the existence of massless charged particles is subject to a, still missing, complete solution of the problem of {\it collinear} infrared divergencies in quantum field theory, while in classical field theory their possible existence relies on the existence of a consistent dynamics including {\it radiation reaction}, {\it i.e.} the self-interaction of the particle caused by emission of radiation. For an analysis of the delicate interrelation between classical and quantum aspects of the radiation problem for {\it massive} charges see {\it e.g.} \cite{HM1}-\cite{IT2}.

The present paper faces the problem of the construction of a consistent classical electrodynamics of massless charged particles. For a massive particle the ``solution'' of this problem amounts to postulate $ i)$ that the field generated by the particle satisfies Maxwell's equations, and $ii)$ that the equation of motion of the particle is the Lorentz-Dirac equation \eref{ldf}. Since this system of equations can not be derived from an action, and so the N\"other procedure is not available, the construction of a - in the sense of distributions - conserved and
well-defined total energy-momentum tensor is a delicate and non-trivial issue \cite{R,LM}. Nevertheless this construction is of fundamental importance  since - ultimately - it is precisely local four-momentum conservation that imposes the Lorentz-Dirac equation. Correspondingly we will consider this conservation paradigm as fundamental also for the dynamics of massless charges.

In absence of an action principle from which to derive the theory - {\it i.e.} the equations of motion and the conservation laws -   we base our strategy to construct a consistent dynamics of massless charged particles on the following principles:
\begin{itemize}
\item Maxwell's equations;
\item relativistic invariance and reparameterization invariance of the lightlike trajectory;
\item local four-momentum conservation.
\end{itemize}
Notice in particular that no {\it a priori} assumption will be made about the equation of motion of the particle. Our starting point will be the, only recently derived, exact expression of the electromagnetic field generated by a massless charged particle in generic motion \cite{AL1,AL2}.  While for a massive particle at fixed time the field diverges only on the particle's position, for a massless particle the field diverges on a {\it string} ending at the particle's position and is, thus, profoundly more singular.

Since local conservation of four-momentum is one of our primary concerns, it is indispensable to construct a well-defined, possibly conserved, energy-momentum tensor: this is a  non-trivial task since the pronounced singularities present in the electromagnetic field turn the {\it formal} energy-momentum tensor  \eref{t0} into an ill-defined object, that is not a distribution. The construction of a
{\it renormalized}, {\it i.e.} in the distributional sense well-defined energy-momentum of the electromagnetic field, is a crucial achievement of the present paper: once such a tensor has been constructed  - we add, in a unique way - the equation of motion of the particle can indeed be {\it derived} requiring  conservation of the total (field + particle) energy-momentum tensor.

The main results of this paper are a) that the renormalized energy-momentum tensor of the electromagnetic field is separately conserved, b) that, correspondingly, to this field no radiation is associated and c) that nonetheless the dynamics of a system of interacting massless charges is perfectly consistent and in agreement with four-momentum conservation. Indeed, as a consequence of b) the equation of motion of a massless charged particle is the ``standard'' Lorentz-equation, without any self-force. In other words: if the total four-momentum must be locally conserved, a massless charged particles does experience no radiation reaction and does not emit {\it bremsstrahlung}.

These results provide in particular the proof of the claim made in \cite{K} that massless charges do not radiate. The preliminary analysis of \cite{K} is, however, based on an electromagnetic field that does {\it not} satisfy Maxwell's equations. Our rigorous analysis clarifies also previous, partially contradictory, attempts to  face the problem of radiation reaction for massless charges \cite{KS,Y}. In a way, still to analyze, we hope that our results might shed also new light on the (possible inconsistency of the) quantum dynamics of such particles.

The paper is written in a self-contained way, being organized as follows. In Section 2, starting from the formal (singular) energy-momentum tensor of the electromagnetic field produced by a - massive or massless - point-like particle, we state four general requirements that must be fulfilled by  the renormalized energy-momentum tensor. Once such a tensor has been constructed the implementation of total local four-momentum conservation leads to a uniquely determined equation of motion for the particle. In Section 3 we illustrate this procedure for a massive particle, retrieving the standard Lorentz-Dirac equation. In Section 4 we present the peculiar features of the electromagnetic field produced by a massless particle, distinguishing  {\it bounded} and {\it unbounded} trajectories. In Section 5 we introduce a Lorentz-invariant regularization of this field and regain a finite {\it putative} self-force, previously known in the literature \cite{KS}, that diverges however for uniform motions. As long as we insist on a {\it regularity paradigm} specified in Section \ref{uniq1} - that is essentially equivalent to the fact that the four-momentum of the electromagnetic field of a particle in uniform motion is finite - according to our framework this self-force does not play any role in the electrodynamics of massless charges. If, on the other hand, we renounce to this paradigm, the construction of a conserved total energy-momentum tensor in presence of this self-force seems rather difficult, if not impossible, but remains in principle an open question. This issue will be addressed in the concluding Section 10, while in the rest of the paper we will insist on our regularity paradigm.
In Section 6 we construct the renormalized energy-momentum tensor for massless particles and in Section 7 we derive its uniqueness and conservation properties and show, in particular, that massless charges do not emit radiation. In Section 8 we show explicitly that for {\it unbounded} trajectories the total four-momentum of the electromagnetic field of a massless charge is finite and conserved - actually vanishing - in agreement with the results of Section 7. In Section 9 we derive eventually the dynamics of a massless particle in presence of an external field and the dynamics of a system of two massless particles, finding that the equations of motion driving these systems are mathematically perfectly consistent. Section 10 contains a summary and a discussion of open problems, especially  the role of the self-force of ref. \cite{KS} and the relation between classical and quantum theories of massless charges. More involved proofs and computations are relegated to four appendices.

\section{Point-particles and singular energy-momentum tensors}

We begin the paper presenting a slight generalization of the procedure employed in \cite{LM} to face the radiation reaction problem, or equivalently the self-interaction problem, of charged point-particles. The procedure we propose entails universality character in that it admits, conceptually immediate, extensions to the radiation reaction problem of extended charged objects, {\it i.e.} p-{\it branes} \cite{KL}. As observed in the Introduction, the self-interaction is in general a {\it non-lagrangian} type of interaction - it can not be derived from an action  - and so we consider as an alternative fundamental principle {\it four-momentum conservation}.

We parameterize the particle's world-line $\gamma$ through the four $C^\infty$-functions $y^\m(\la)$ and indicate its four-velocity and four-acceleration respectively with $u^\m=dy^\m/d\la$ and  $w^\m=du^\m/d\la$.
We denote the spatial velocity and acceleration respectively with
$\vec v=\vec u/u^0=d\vec y/dt$ and $\vec a=d\vec v/dt$. We consider $\la$ as a generic parameter, in general not identified with proper time, so that our formalism applies equally well to massive ($u^2>0$) and massless ($u^2=0$) particles. This means that all observable quantities must be invariant under a reparameterization of the world-line
\be\label{repara}
\la\ra\la'(\la), \quad\quad y'^\m(\la')=y^\m(\la).
\ee
The electromagnetic field generated by a particle with charge $e$ must satisfy Maxwell's equations - in the distributional sense -
\begin{equation}
\partial_{\mu}F^{\mu\nu}= e\!\int\!\delta^{4}(x-y(\lambda))\,dy^\n\equiv j^{\n},\quad \quad\partial_{[\m}F_{\n\rho]}=0.\label{Max}
\end{equation}

\subsection{Renormalized energy-momentum tensor: general construction}\label{cot}

Own to the point-nature of the particles the  solutions of equations \eref{Max} - the Li\'enard-Wiechert field \eref{lw} for massive charges and the field \eref{unbound} for massless ones - in general diverge on a {\it singularity-locus}: for a massive particle this locus is the world-line $\g$, while for a  massless particle it is a two-dimensional surface $\G$,  whose boundary (in the case of a bounded trajectory) is $\g$, see Section \ref{bau}. As shown in Section \ref{bau}, for an unbounded trajectory the surface $\G$ acquires an additional boundary, having the topology of a strip. In the complement of the singularity-locus in $\mathbb{R}^4$ the fields are, actually, of class $C^\infty$.

Although the electromagnetic field by definition is a (tempered) distribution,
the {\it formal} energy-momentum tensor
\be\label{t0}
\Theta^{\a\bt}=(F|F)^{\a\bt}
\ee
is not. Given two antisymmetric tensors $F^{\m\n}$ and $G^{\m\n}$ we use the shorthand notation
\be\label{fg}
(F|G)^{\a\bt}=  F^{\m(\a}\, G^{\bt)}{}_\m + \frac{1} {4}\,
\eta^{\a\bt} F^{\m\n} G_{\m\n}.
 \ee
A product of distributions is in general, in fact, not a distribution. For a massive particle, for example, near the world-line $\Theta^{\a\bt}$
diverges as $1/r^4$ and it is thus locally non-integrable. This circumstance has two dramatic consequences: I) the four-momentum volume integrals at fixed time $P^\bt_V= \int_V \Theta^{0\bt}(t,\vec x)\, d^3x$ are divergent if $V$ contains the particle, and II)  the distributional four-divergence $\pa_\a \Theta^{\a\bt}$ is ill-defined.

Before one can face the problem of four-momentum conservation one must thus first of all construct a mathematically well-defined energy-momentum tensor, that in particular admits derivatives. More precisely one must construct a {\it renormalized} energy-momentum tensor $T^{\a\bt}_{em}$ of the electromagnetic field that is a {\it distribution}, {\it i.e.} that belongs to the dual ${\cal S}^\prime({\mathbb R}^4)$ of the Schwartz space of test functions ${\cal S}({\mathbb R}^4)$. On general grounds we impose on this tensor the four basic requirements (see also \cite{R}):
\begin{itemize}
\item[1)] $T^{\a\bt}_{em}$ is a distribution;
\item[2)] $T^{\a\bt}_{em}$ is covariant, symmetric, traceless and  reparameterization-invariant, as is $\Theta^{\a\bt}$;
\item[3)] $T^{\a\bt}_{em}(x)= \Theta^{\a\bt}(x)$ for every $x$ in the complement of the singularity-locus;
\item[4)] the four-divergence of  $T^{\a\bt}_{em}$ is {\it multiplicatively} supported on $\g$, {\it i.e.}
\be
\label{dtloc}\pa_\a T^{\a\bt}_{em}= -\int \!f^\bt(\la)\, \dl^4(x-y(\la))\,d\la
\ee
for some four-vector $f^\bt(\la)$ ``multiplying'' the $\dl$-function.
\end{itemize}

The physical interpretation of the first three requirements is self-evident.
Requirement 3) implies in particular the peculiar feature that the renormalized tensor $T^{\a\bt}_{em}$  is {\it determined only modulo terms  supported on the singularity-locus} - an {\it intrinsic} ambiguity that will play an important role in the following, see also \cite{KL}.

The origin of requirement 4) is local four-momentum conservation.
Introduce the total energy-momentum tensor as
\be\label{ttot}
T^{\a\bt}=T^{\a\bt}_{em}+ T^{\a\bt}_p, \quad\quad  T^{\a\bt}_p=\int \!u^\a p^\bt \dl^4(x-y(\la))\,d\la,
\ee
$ T^{\a\bt}_p$ being the standard contribution of the particle. For a massive particle we have $p^\bt= {mu^\bt}/{\sqrt{u^2}}$, while for a massless one we have $p^\bt=gu^\bt$, where $g(\la)$ is the {\it einbein}-field  ensuring reparameterization invariance;
under a reparameterization \eref{repara} it transforms as
\[
g'(\la')=\frac{d\la'}{d\la} \,g(\la).
\] If  \eref{dtloc} holds, enforcing local total four-momentum conservation
\be\label{dttt}
\pa_\a T^{\a\bt}= \int\!\left(\frac{dp^\bt}{d\la}-f^\bt(\la)\!\right)\dl^4(x-y(\la))\,d\la=0,
\ee
one derives the ``Lorentz-equation'' of motion for the charge
\be\label{eqm}
\frac{dp^\m}{d\la}=f^\m,
\ee
that - in absence of external fields - identifies $f^\m$ as the (automatically finite) {\it self-force}. Consistency of this equation requires then, further, that this force obeys the identity $u_\m f^\m=0$.

There is a second, related, reason for insisting on requirement 4), which is more directly tied to {\it covariance}. To explain it we recall a known basic fact about energy-momentum tensors. If  $T^{\a\bt}_{em}$ is a generic tensor satisfying  $\pa_\a T^{\a\bt}_{em} =0$, then (under certain regularity conditions at spatial infinity) the formal integrals $P^\bt_{em}=\int T^{0\bt}_{em}\, d^3x$ - apart from being conserved - form a {\it four-vector},  see {\it e.g.} \cite{W}. If, on the contrary,  we know only that $T^{\a\bt}_{em}$ is a tensor, the four quantities $P^\bt_{em}$,
apart from not being conserved, in general do not transform covariantly under Lorentz-transformations. In the case of a single particle we want the total four-momentum $P^\bt_{em}+p^\bt$ to be conserved and {\it covariant} and so, since the four-momentum $p^\bt$ of the particle is a four-vector, $P^\bt_{em}$ must be a four-vector, too - although in general obviously $\pa_\a T^{\a\bt}_{em}\neq 0$. However, if $\pa_\a T^{\a\bt}_{em}$ has the particular form \eref{dtloc}, integrating this equation over whole space and over the time interval $(-\infty,t]$ (and assuming appropriate asymptotic behaviors for $T^{\a\bt}_{em}$) we derive the explicit expression
\be
P^\bt_{em}(t)= \int\! T^{0\bt}_{em}(t,\vec x)\,d^3x=-\int_{-\infty}^{\la(t)} \!f^\bt(\la)\,d\la.
\ee
If one regards $t$ as a function of the invariant parameter $\la$, then these integrals form, indeed, a four-vector.

Were the four-divergence of $T^{\a\bt}_{em}$ supported on $\g$ in a {\it non-multiplicative} way, {\it i.e.} would the vector $f^\bt(\la)$ in \eref{dtloc} be replaced by a derivative operator acting on the $\dl$-function, {\it e.g.} $f^\bt(\la) \ra h^{\bt\m}(\la)\pa_\m$, then both above properties would fail:  $P^\bt_{em}$ would not be a four-vector and there would exist no Lorentz-equation of motion guaranteeing the vanishing of  $\pa_\a T^{\a\bt}$ in \eref{dttt}.

A we will see, the requirements 1)-4), together with the {\it regularity paradigm} introduced in Section \ref{uniq1}, determine $T^{\a\bt}_{em}$ uniquely - in the massive as well as in the massless case - furnishing thus a uniquely determined equation of motion for the particle, taking radiation reaction into account.

\section{Massive charges}\label{mc}

We recall now briefly from \cite{LM} how one can implement the requirements 1)-4) in the case of a massive particle, following a time-like trajectory.

For a time-like trajectory Maxwell's equations \eref{Max} entail the Li\'enard-Wiechert solution (for the moment we ignore the external field)
\be\label{lw}
F^{\m\n}=\frac{e}{4\pi}\bigg(u^{2}\frac{L^{\mu}u^{\nu}}
{\left(uL\right)^{3}}+\frac{L^{\mu}\big((uL)w^{\nu}-(wL)
u^{\nu}\big)}{(uL)^{3}}\bigg) -(\mu\leftrightarrow\nu),
\ee
where, we recall, $u^\m$ and $w^\m$ indicate the four-velocity and four-acceleration w.r.t. to a generic parameter $\la$.
For contractions we use the notation $(uL)=u^\m L_\m$ {\it etc}. and we have set \be\label{lm}
L^\m=x^\m-y^\m(\la).
\ee
All kinematical variables in \eref{lw} are evaluated at the retarded parameter $\la(x)$ defined by the conditions
\be\label{rt}
L^2=(x-y(\la))^2=0,\quad\quad x^0-y^0(\la)>0.
\ee

The, conceptually simple, strategy to determine a tensor $T^{\a\bt}_{em}$ satisfying the requirements 1)-4) developed in \cite{LM} proceeds as follows. Introduce a regularized field $F^{\m\n}_\ve(x)$, that is obtained from \eref{lw} by replacing $\la(x)$  with the regularized  retarded parameter $\la_\ve(x)$, determined by the conditions
\be\label{rte}
L^2=(x-y(\la))^2=\ve^2,\quad\quad x^0-y^0(\la)>0,
\ee
where $\ve>0$ is a regulator with the dimension of length.  The field  $F^{\m\n}_\ve$ can be seen to be a $C^\infty$-distribution and in particular one has the distributional limit
\be\label{sfe}
{\cal S}^\prime \!- \!\lim_{\ve\ra 0} F^{\m\n}_\ve = F^{\m\n}.
\ee
However, the {\it regularized} energy-momentum tensor
  \be
 \label{treg}
\Theta^{\a\bt}_\ve= (F_\ve| F_\ve)^{\a\bt}
 \ee
does not admit a distributional limit as $\ve\ra 0$.  Before taking this limit
one must identify - and subtract - the singular part $\Theta^{\a\bt}_\ve{}\big|_{div}$ of this tensor,   a divergent ``counterterm'',  that in line with requirement 3) must be supported on $\g$. More precisely, the renormalized energy-momentum tensor of the electromagnetic field is the distributional limit
 \be
\label{lim0}
T^{\a\bt}_{em} ={\cal S}^\prime \!- \!\lim_{\ve\ra 0}
\left(\Theta^{\a\bt}_\ve -\Theta^{\a\bt}_\ve{}\big|_{div}\right)
\equiv {\cal S}^\prime \!- \!\lim_{\ve\ra 0}
\bigg(\Theta^{\a\bt}_\ve - \frac{e^2}{32\ve} \int\! \bigg(\frac{u^\a u^\bt}{u^2} -
\frac{1} {4} \,\eta^{\a\bt}\bigg)\dl^4 (x-y(\la))\sqrt{u^2}\,d\la\bigg).
 \ee
In \cite{LM} it has indeed been proven  $i)$ that the limit \eref{lim0} exists, so that $T^{\a\bt}_{em}$ is a distribution,  and $ii)$ that the four-divergence of the so defined energy-momentum tensor is given by
\be\label{div0}
 \pa_\a T^{\a\bt}_{em}=- \frac{e^2}
{6\pi} \int\! \left(\frac{dW^\bt}{d\la} +W^2
u^\bt\right)\delta^4(x-y(\la))\,d\la,
\ee
where $W^\m$ is the reparameterization invariant four-acceleration
\be
W^\m =\frac{d^2y^\m}{ds^2}, \quad\quad \frac{d}{ds}= \frac1{\sqrt{u^2}}\frac{d}{d\la}.
 \ee
Notice that the tensor  $\Theta^{\a\bt}_\ve{}\big|_{div}$, diverging as $1/\ve$, is manifestly invariant under a reparameterization \eref{repara}. The tensor \eref{lim0} satisfies thus the requirements 1)-4).

\subsection{Lorentz-Dirac equation and external field}\label{uad}

According to the general strategy represented by equations \eref{dtloc}-\eref{eqm}, from \eref{div0}  we deduce that local four-momentum conservation, {\it i.e.} $ \pa_\a T^{\a\bt}=0$, forces  the particle to satisfy the celebrated Lorentz-Dirac equation \be\label{LD}
\frac{dp^\m}{d\la}= \frac{e^2}{6\pi} \left(\frac{dW^\m}{d\la} +W^2
u^\m\right).
\ee

In presence of an external field ${\cal F}^{\m\n}$, satisfying the homogeneous Maxwell equations
\be\label{maxf}
\partial_{\mu}{\cal F}^{\mu\nu}= 0=\partial_{[\m}{\cal F}_{\n\rho]}=0,
\ee
the formal  energy-momentum tensor of the total field is
\be\nn
\Theta^{\a\bt} =(F+{\cal F}|F+{\cal F})^{\a\bt},
\ee
and the renormalized energy-momentum tensor, satisfying 1)-4), is given by
\be\label{temf}
\mathbb{T}^{\a\bt}_{em}={T}^{\a\bt}_{em}+2(F|{\cal F})^{\a\bt}+ ({\cal F}|{\cal F})^{\a\bt},
\ee
with ${T}^{\a\bt}_{em}$ still given in \eref{lim0}.
Assuming, in fact, that ${\cal F}^{\m\n}$ is a (regular) $C^\infty$-distribution, the new terms appearing in \eref{temf} w.r.t. \eref{lim0} are distributions, so that no new counterterms are needed.

Using that for generic antisymmetric fields $F^{\m\n}$  and $G^{\m\n}$, obeying the Bianchi identities $\partial_{[\m}{ F}_{\n\rho]}=0=\partial_{[\m}{ G}_{\n\rho]}$, the tensor \eref{fg} satisfies the Leibnitz-rule
\be\label{rule}
\pa_\a (F|G)^{\a\bt}=\frac12\left(\pa_\a F^{\a\mu}G_\m{}^\bt+\pa_\a G^{\a\mu}F_\m{}^\bt\right)\!,
\ee
in virtue of \eref{Max} and \eref{maxf}, from  \eref{temf} instead of \eref{div0}
we obtain now the identity
\be\label{id1}
 \pa_\a \mathbb{T}^{\a\bt}_{em}=-
 \int\!\left(\frac{e^2}
{6\pi}\left(\frac{dW^\bt}{d\la} +W^2
u^\bt\right)\!+e{\cal F}^{\bt\n}u_\n\right) \delta^4(x-y(\la))\,d\la.
\ee
In this way, imposing that the tensor $T^{\a\bt}=\mathbb{T}^{\a\bt}_{em}+ T^{\a\bt}_p$ has vanishing four-divergence, one derives the equation of motion
\be\label{ldf}
\frac{dp^\m}{d\la}= \frac{e^2}{6\pi} \left(\frac{dW^\m}{d\la} +W^2
u^\m\right)\!+e{\cal F}^{\m\n}u_\n,
\ee
whose r.h.s. satisfies indeed $u_\m f^\m=0$.

\subsubsection{Total four-momentum}\label{tfm}

If we impose that the external field ${\cal F}^{\a\bt}$ is at any instant of compact spatial support and that the particle follows an unbounded trajectory - for which the acceleration $\vec a(t)$ for $t\ra-\infty$ vanishes sufficiently fast - the total four-momentum $P^\bt=\int T^{0\bt} d^3x$ of the system is finite, as well as conserved. From the equations above we can also derive an explicit expression for it.

The energy-momentum tensor of the external field $({\cal F}|{\cal F})^{\a\bt}$ in \eref{temf} is separately divergenceless and so its four-momentum $P^\bt_{ext}=\int ({\cal F}|{\cal F})^{0\bt}d^3x$ is separately conserved. Integrating equation \eref{id1} over whole space and applying Gauss' theorem, using that at spatial infinity all fields vanish, one finds an equation for the time-derivative of the electromagnetic four-momentum $P^\bt_{em}(t)=\int \mathbb{T}^{0\bt}_{em}\,d^3x$.
Since for $t\ra-\infty$ the acceleration vanishes sufficiently fast we have
$\lim_{t\ra-\infty} P^\bt_{em}(t)= P^\bt_{ext}$, so that the so obtained equation determines $P^\bt_{em}(t)$ uniquely. The result for the total four-momentum  $P^\bt= p^\bt(t)+ P^\bt_{em}(t)$ reads eventually
\be\label{ptotm}
 P^\bt = p^\bt(t)- \frac{e^2}{6\pi}\,\bigg(W^\bt(t)+\int_{-\infty}^{\la(t)}W^2u^\bt\,d\la\bigg)-e
 \int_{-\infty}^{\la(t)}{\cal F}^{\bt\n}u_\n\,d\la+P^\bt_{ext}.
\ee
The first term is the four-momentum of the particle, the second represents the emitted radiation, the third the interference between the Li\'enard-Wiechert and the external field and the fourth term is the constant four-momentum of the external field. Notice that \eref{ptotm} is conserved thanks to \eref{ldf}.

\subsection{Uniqueness, finite counterterms and a regularity paradigm}\label{uniq1}

From a conceptual point of view the - otherwise stringent - derivation of the Lorentz-Dirac equation presented above is tightly related to the uniqueness of an energy-momentum tensor $T^{\a\bt}_{em}$ satisfying the requirements 1)-4). These conditions are ``solved'' by the expression \eref{lim0} but, as observed previously, condition 3) introduces an indeterminacy consisting in the freedom to add to  \eref{lim0} a {\it finite} counterterm $D^{\a\bt}$. To preserve conditions 1)-4) this tensor is subject to the constraints:
\begin{itemize}
\item[a)] $ D^{\a\bt}$ must be a distribution supported on $\g$, with the dimension of an energy density;
\item[b)] $ D^{\a\bt}$ must be covariant, symmetric,  traceless and reparameterization invariant;
\item[c)] the four-divergence $\pa_\a D^{\a\bt}$ must be multiplicatively supported on $\g$.
\end{itemize}
For a spinless  particle this tensor must be constructed with $\pa^\m$, $u^\m$, $w^\m$ and the successive derivatives of the world-line, while $y^\m$ itself would violate translation invariance. The most general form of a tensor $D^{\a\bt}$ satisfying a) and b) is
 \be\label{finc}
D^{\a\bt}= {e^2}\!\int\! \left(c_1\,u^{(\a} W^{\bt)}+c_2\,u^{(\a} \pa^{\bt)}\right)\!\dl^4 (x-y(\la))\,d\la,
\ee
where $c_1$ and $c_2$ are dimensionless numerical coefficients. To impose property c) we compute
\[
\pa_\a D^{\a\bt}= \frac{e^2}2\!\int
\!\left(c_1\frac{dW^\bt}{d\la}
+u^\bt\big(c_1W^\a\pa_\a+c_2\,\square\big)\!\right)
\delta^4(x-y(\la))d\la.\label{divb}
\]
As one sees, the two terms multiplying $u^\bt$ are {\it non-multiplicatively} supported on $\g$, unless $c_1=c_2=0$. This means that no finite counterterms are available.

What we have just shown is that the energy-momentum tensor of the electromagnetic field \eref{lim0}, compatible with total four-momentum conservation, is uniquely determined.
With this respect we must add that in establishing  the general form \eref{finc} we have implicitly forbidden the presence of - covariant and dimensionally correct - {\it singular} contributions to  $D^{\a\bt}$, as {\it e.g.}
\[D_{(sing)}^{\a\bt}=
{e^2}\!\int\!\frac{1}{W^2} \,u^{(\a} \pa^{\bt)}\,
\square\,\dl^4 (x-y(\la))d\la.
 \]
Such terms diverge for particles in uniform motion, for which $W^\m=0$, and consequently also the four-momentum integrals for free particles would be divergent, a behavior that we consider as unphysical. Correspondingly here, and henceforth also for massless particles, we adopt what we call a {\it regularity paradigm}, according to which all local counterterms - as the original energy-momentum tensor - must admit finite limits for uniform motions. This excludes in particular terms like $D_{(sing)}^{\a\bt}$ and all similar ones having (powers of) the acceleration at the denominator.

As anticipated in the Introduction, since this paradigm might not be accepted by all theoreticians, in Section \ref{sao} we  analyze the kind of electrodynamics that might emerge if one renounces to this paradigm.

\subsection{Heuristic arguments for the self-force}

In the literature there exists a variety of {\it heuristic} derivations of the self-force, {\it i.e.} the four-vector at the r.h.s. of \eref{LD}, but eventually the Lorentz-Dirac equation must be {\it postulated}.

One such derivation starts from the  relativistic Larmor formula $dP_{rad}^\m/d\la=(e^2/6\pi)W^2u^\m $ - representing the radiated four-momentum that reaches infinity - and ends adding ``by hand'' the {\it Schottky term} $(e^2/6\pi)dW^\m/d\la$, to cope with the identity $u_\m f^\m=0$.

An apparently more systematic procedure consists in considering the regularized field $F^{\m\n}_\ve(x)$ defined above, or to resort to some other regularization,
and to introduce the regularized self-force
\be\label{self00}
 f^\m_\ve\equiv eF^{\m\n}_\ve(y)u_\n.
\ee
Expanding it around $\ve=0$ one obtains
\be\label{self0}
f^\m_\ve=-\frac{e^2\sqrt{u^2}}{8\pi \ve}\, W^\m+\frac{e^2}{6\pi} \left(\frac{dW^\m}{d\la} +W^2
u^\m\right)\!+o(\ve).
\ee
If one subtracts the divergent term proportional to $1/\ve$ - invoking some kind of mass renormalization - the finite terms as $\ve\ra 0$ reproduce then indeed equation \eref{LD}.

Apart from the ``invoked'' result one must however keep in mind that  {\it a priori} all these procedures have no ``fundamental'' basis and that the recovery of \eref{LD} - the same equation that above has been derived enforcing (the fundamental requirement of) four-momentum conservation - has to be considered merely as a,  still not fully understood, {\it coincidence}.

\section{Massless charges and their electromagnetic field}

Massless charges follow lightlike trajectories so that the four-velocity satisfies
\[
u^2=0.
\]
The electromagnetic field generated by such particles has been determined in \cite{AL1,AL2}
and reveals several unexpected features, according to whether the trajectory is bounded or unbounded. Below we summarize the most important ones.

\subsection{Singularity surface}

The main differences between the field generated by a massless charge and the Li\'enard-Wiechert field \eref{lw} arise from the peculiar singularity-locus of the former: at fixed time it is a {\it string} whose one endpoint is the particle's position, rather than solely the particle's position itself. During time evolution this string sweeps out a surface $\Gamma$ parameterized by
\be\label{Ga}
\G^\m(\la,b)=y^\m(\la)+bu^\m(\la), \quad\quad b\ge0.
\ee
Notice that $\G^\m(\la,0)=y^\m(\la)$, {\it i.e.}  the boundary of $\G$ {\it includes} the world-line $\gamma$. As we will see below, for unbounded trajectories the boundary of $\G$ acquires indeed an additional curve, arising from $b\ra\infty$.
For a lightlike trajectory the reparameterization invariant proper time $ds=\sqrt{u^2}\,d\la$ is not available, but nonetheless all physical observables must be independent of the way one parameterizes the world-line $y^\m(\la)$. Since the surface $\Gamma$ is such an observable, under a reparameterization \eref{repara} the variable $b$ must transform according to
\be\label{brep}
b\ra b'=\frac{d\la'}{d\la}\,b.
\ee
From the spatial components of \eref{Ga}, parameterizing the world-line with time, $\la=y^0(\la)$, and setting $ \G^0=\la+b\equiv t$, one sees that the singularity string attached to the particle at time $t$ is given by
\be\label{ga}
\vec\G(t,b)=\vec y(t-b)+b\vec v(t-b),\quad\quad b\ge0.
\ee

\subsubsection{Bounded and unbounded trajectories}\label{bau}

For a bounded trajectory we have $|\vec y(t)|<M$ for all $t$ for some $M$, and in this case from \eref{Ga}, or \eref{ga}, it follows that as $b\ra\infty$, all points of the singularity surface tend to infinity and so no additional boundary  of $\G$ arises.

As ``unbounded trajectories'' we consider motions that in the infinite past approach sufficiently fast a straight line
\be\label{asym}
\vec y(t)\ra \vec v_\infty t,\quad \quad t\ra -\infty,
\ee
where the constant asymptotic velocity  satisfies obviously $|\vec v_\infty|=1$.
In this case for $b\ra \infty$ \eref{ga} gives
\be
\lim_{b\ra\infty}\vec\G(t,b)=\lim_{b\ra\infty}\left(\vec v_\infty(t-b)+b\vec v_\infty\right)=\vec v_\infty t,
\ee
meaning that $\G$ acquires as additional boundary a virtual  world-line ${\cal L}$, parameterized  by
\be\label{virt}
y^\m_{\cal L}(\la)=(1,\vec v_\infty)\,\la,
\ee
that corresponds to a {\it fictitious} particle in strictly linear motion with constant velocity $\vec v_\infty$. For unbounded trajectories the boundary of $\G$ is thus
$$
\pa \G=\g\cup{\cal L},
$$
so that the singularity string \eref{ga} for every $t$ has a {\it finite} extension: it starts from the particle's position and ends on a point of ${\cal L}$.

\subsection{The field}

For a {\it bounded} trajectory the retarded electromagnetic field satisfying Maxwell's equations \eref{Max} is given by the distribution (disregarding the external field)
\begin{align}
F^{\m\n}&=\frac{e}{4\pi}\,{\cal P}\, \bigg(\frac{L^{\mu}\big((uL)w^{\nu}-(wL)
u^{\nu}\big)}{(uL)^{3}} -(\mu\leftrightarrow\nu)\!\bigg)
 +\frac{e}{2}\!\int\! b\,\big(u^{\mu}w^{\nu}-u^{\nu}w^{\mu}\big)
\delta^{4}\big(x-\Gamma(\lambda,b)\big)db d\lambda\nn\\
&\equiv F^{\m\n}_{reg}+F^{\m\n}_{sing},
\label{bound}
\end{align}
where ${\cal P}$ indicates the ``principal part'' of the expression between parentheses, see  \cite{AL2}. In the regular field $F^{\m\n}_{reg}$ the kinematical variables are evaluated at the retarded time defined in \eref{rt}, while in the singular field $F^{\m\n}_{sing}$ - supported on $\G$ - they are evaluated at the integration variable $\la$. In accordance with \eref{Ga} it is understood that the integration region for $b$ is restricted to the {\it positive} real axis - a convention that we will maintain for all $b$-integrals of this paper. Notice that both fields in \eref{bound} are reparameterization invariant, see in particular \eref{brep}.

The field \eref{bound} is the distributional limit under $u^2\ra 0$
of the field \eref{lw}: the distributional limit of the Coulomb field - the first term of \eref{lw}, proportional to $u^2$ - is actually zero, coinciding thus with its point-wise limit. The distributional limit of the radiation field - the second term of \eref{lw}, proportional to $w$ - produces instead the sum $F^{\m\n}_{reg}+F^{\m\n}_{sing}$. A detailed analysis reveals in particular that the electric flux  is distributed with equal weights between these two fields:
\be\label{flux1}
\partial_{\mu}F^{\mu\nu}_{reg}=\frac 12\,j^\n=\partial_{\mu}F^{\mu\nu}_{sing}.
\ee

For an {\it unbounded} trajectory (as specified above) the electromagnetic field acquires an additional contribution, the solution of Maxwell's equations being now
\be\label{unbound}
F^{\m\n}=F^{\m\n}_{reg}+F^{\m\n}_{sing}+F^{\m\n}_{sw},
\ee
where the new term
\be\label{fsw}
F^{\m\n}_{sw}=
\frac{e}{2\pi}\frac{v^{\mu}_\infty x^{\nu}-v^{\nu}_\infty x^{\mu}}{x^{2}}\,\delta (v_\infty x), \quad\quad v^\m_\infty=(1,\vec v_\infty),
\ee
is a {\it shock-wave} produced by a  virtual charged particle traveling on the straight line ${\cal L}$ \eref{virt}. For such trajectories the electric flux is distributed according to the equations (replacing \eref{flux1})
\begin{align}
\label{flux2}
\partial_{\mu}F^{\mu\nu}_{reg}=\frac 12\,\big(j^\n-j^\n_{\cal L}\big)=\partial_{\mu}F^{\mu\nu}_{sing},\quad\quad
\pa_\m F^{\m\n}_{sw}=j^\n_{\cal L},
\end{align}
where the current producing  the shock-wave field is
\be\label{jl}
j_{\mathcal L}^\m(x)=e\!\int\! v^\m_\infty\, \dl^4(x-\la v_\infty)\,d\la.
\ee

\section{Regularization}\label{regul}

The formal energy-momentum tensor \eref{t0} of the fields \eref{bound} and \eref{unbound} is ill-defined, in that the product of distributions in general is not a distribution.
To carry out the construction of Section \ref{cot} in the case of lightlike trajectories we resort to the same regularization employed in Section \ref{mc}. In what follows we will restrict ourselves to bounded trajectories, presenting the variant for unbounded ones in Section \ref{ut}.

For bounded trajectories we introduce the regularized field
\be\label{freg}
F^{\m\n}_\ve(x)=\frac{e}{4\pi}\,\frac{L^{\mu}\big((uL)w^{\nu}-(wL)
u^{\nu}\big)}{(uL)^{3}}\bigg|_{\la_\ve(x)} -(\mu\leftrightarrow\nu),
\ee
where the kinematical variables are evaluated at the regularized retarded parameter $\la_\ve(x)$ defined in \eref{rte}. It can indeed be shown that, as in the massive case, this field is everywhere regular - more precisely of class $C^\infty$ - and that it admits the distributional limit \cite{AL2}
\be\label{limf}
{\cal S}^\prime \!- \!\lim_{\ve\ra 0} F^{\m\n}_\ve = F^{\m\n}_{reg}+F^{\m\n}_{sing},
 \ee
{\it i.e.} precisely the field \eref{bound}. A further, conceptual as well as technical, advantage of this regularization is its {\it manifest} Lorentz-invariance.

\subsection{Self-force from a heuristic argument}\label{self}

Before proceeding with the construction of the renormalized energy-momentum tensor we derive, in analogy with \eref{self00} and \eref{self0}, a ``putative'' self-force. The formal self-force
$eF^{\m\n}(y)u_\n$ is again divergent - $F^{\m\n}(x)$ diverges on $\G$ and {\it a fortiori} it diverges on the world-line - and so in analogy with \eref{self00} we regularize it according to
$eF^{\m\n}(y)u_\n\ra  eF^{\m\n}_\ve(y)u_\n$.

The so regularized self-force is again Lorentz- and reparameterization-invariant and, as long as $\ve>0$, it is finite for every $\la$.
Its expansion for $\ve\ra 0$ is a bit cumbersome, although conceptually simple, and we perform it in Appendix \ref{sef}. To keep track of reparameterization invariance in a manifest way it is convenient to introduce the {\it reparameterization-invariant} derivative
\be\label{ds}
\frac{d}{d\s}\equiv \frac{1}{(-w^2)^{1/4}}\frac{d}{d\la}, \quad\quad w'^2(\la')= \bigg(\frac{d\la}{d\la'}\bigg)^4 w^2(\la),
\ee
where the parameter $\s$ resembles in a certain sense the proper time of a massive particle. We denote the related four-velocity by $U^\m=dy^\m/d\s$.

Introducing the reparameterization invariant quantities
\be\label{ymn}
y^{N\m}=\frac{d^Ny^\m}{d\s^N},\quad\quad y^{MN}=y^{M\m}y^N{}_\m,
\ee
in Appendix  \ref{sef} we derive the Laurent-type expansion of the self-force relative to the parameter $\s$ (to be compared with \eref{self0})
\be\label{sf}
f^\m_\ve\equiv eF^{\m\n}_\ve(y)U_\n= \frac{e^2}{4\pi} \left(\frac{1}{\ve^{3/2}}\,f_{3}^\m+
\frac{1}{\ve}\,f_2^\m+\frac{1}{\ve^{1/2}}\,f_1^\m+f_0^\m\right)+ o\big(\ve^{1/2}\big),
\ee
where
\begin{align}
f_3^\m&=\frac{3}{12^{3/4}}\,y^{2\m},\\
f_2^\m&=-\frac{6}{5\sqrt{12}}\,y^{15}y^{1\mu},\\
f_1^\m&=\frac{3}{4\cdot12^{1/4}}\left(\frac{11}{5}\,y^{34}y^{1\m}+\frac{11}{10}\,
y^{15}y^{2\m}-y^{4\m}\right)\!,\\
f_0^\m&=\frac{2}{5}\left(\!\left(\frac{2}{5}\,\big(y^{15}\big)^2-\frac97\,y^{44}
-\frac{11}{7}\,y^{35}\right)  y^{1\m}-3y^{34}y^{2\m}-y^{15}y^{3\m}+y^{5\m}\right)\!.\label{f0}
\end{align}
A different regularization procedure has been adopted in \cite{KS} - where an  expansion like \eref{sf} has been actually performed for the first time -  and, rather surprisingly, all our forces $f^{\m}_n$ match exactly with those obtained in \cite{KS} apart from, obviously, the overall (divergent) coefficients. In particular the finite self-force $f_0^\m$ is the same, including the overall coefficient $2/5$. The expansion performed by us - based on  \eref{rte} - and the one adopted in \cite{KS} correspond  to replace the Green-function $H(x^0)\dl(x^2)/2\pi$ of the d'Alembertian  respectively with
\[
\frac{H(x^0)}{2\pi}\,\dl(x^2-\ve^2)\quad \mbox{and} \quad \frac{H(x^0)}{2\pi}\,H(x^2)\frac{e^{-x^2/\ve^2}}{\ve^2}.
\]

The fact that the result \eref{sf} is the same might confer in particular to the finite self-force $f^\m_0$ universality character - despite its conflict with causality due to
the presence of the fifth derivative of $y^\m(\la)$, and despite the singularities introduced by the high powers of $w^2$ in the denominator of $y^{N\m}$, {\it i.e.} $1/(-w^2)^{N/4}$. Correspondingly, following \cite{KS}, one could therefore invoke some ``renormalization procedure'' to eliminate the divergent self-forces and {\it postulate} as equation of motion of a massless charge - in presence of an external field ${\cal F}^{\m\n}$ -
\be\label{post}
 \frac{dp^\m}{d\s}=\frac{e^2}{4\pi}\, f^\m_0 +e{\cal F}^{\m\n}U_\n.
\ee\
Notice that by construction the self-force satisfies $U_\m f^\m_0=0$.

Nonetheless we emphasize that also in the present case these derivations entail purely {\it heuristic} character - like \eref{self0} - and that there is no {\it a-priori}-indication  that equation \eref{post} respects four-momentum conservation. As we will show in Section \ref{eom}, as long as we insist on the {\it  regularity paradigm}, this equation is not compatible with four-momentum conservation, since in that case the r.h.s. of \eref{post} does not contain $f^\m_0$. In a certain sense this is obvious since $f_0^\m$ by itself does not satisfy the regularity paradigm, as it diverges as $w^2\rightarrow 0$. As anticipated several times, we will examine the special conditions under which equation \eref{post} might respect four-momentum conservation - upon violating the regularity paradigm - in the concluding section.

\section{Renormalized energy-momentum tensor for massless particles}\label{emt}

We  construct first the renormalized energy-momentum tensor for the field \eref{bound} of a  {\it bounded} trajectory, relegating the modifications needed for unbounded trajectories to Section \ref{ut}.

Since for a massless charge the singularity-locus is the surface $\G$ in \eref{Ga}, requirement 3) of Section \ref{cot} is now specified as
\begin{itemize}
\item[3)] $T^{\a\bt}_{em}(x)= \Theta^{\a\bt}(x)$ for all $x\in{\mathbb R^4}\backslash\G$, \,{\it i.e.}  in the complement of $\G$,
\end{itemize}
where  $\Theta^{\a\bt}$ is given in \eref{t0} and the field is that in \eref{bound}. Accordingly the tensor $T^{\a\bt}_{em}$ is now determined modulo terms supported on  $\G$.

Starting point of our construction are the regularized  $C^\infty$-field \eref{freg} and the related regularized energy-momentum tensor $\Theta^{\a\bt}_\ve$ \eref{treg},  likewise a  $C^\infty$-distribution.  In the complement of $\G$ we have the {\it point-wise} limit
\be\label{pw}
\lim_{\ve\ra0}\Theta^{\a\bt}_\ve(x)= \Theta^{\a\bt}(x),\quad\quad  \forall\, x\in {\mathbb R}^4 \backslash \G.
\ee
Plugging \eref{freg} into \eref{treg} the regularized energy-momentum tensor can be cast in the form (in the following for simplicity we set $e/4\pi=1$)
\be\label{the}
\Theta^{\a\bt}_\ve= A^{\a\bt} + B^{\a\bt}-\frac12 \, \eta^{\a\bt}B^\g{}_\g,
\ee
where
\be\label{ab}
A^{\a\bt}=-\frac{w^2L^\a L^\bt}{(uL)^4}\,,\quad\quad
B^{\a\bt}=\ve^2\left(2\,\frac{(wL)u^{(\a} w^{\bt)}}{(uL)^5}-\frac{w^\a w^\bt}{(uL)^4}-
\frac{(wL)^2u^\a u^\bt}{(uL)^6}\right),
\ee
and it is understood that all kinematical variables are evaluated at the regularized retarded parameter $\la_\ve(x)$ \eref{rte}. Notice that, thanks to $L^2=\ve^2$ and $u^2=0=(uw)$, one has in particular
\be\label{L2e}
A^\g{}_\g=B^\g{}_\g= -\frac{\ve^2 w^2}{(uL)^4},
\ee
in agreement with the tracelessness of $\Theta^{\a\bt}_\ve$.

\subsection{Divergent counterterms and renormalization}

Although the point-wise limit \eref{pw} exists, as in the massive case the {\it distributional} limit ${\cal S}^\prime \!- \!\lim_{\ve\ra 0} \Theta^{\a\bt}_\ve$ does not. Before taking this limit, in analogy with \eref{lim0} we must again identify - and subtract - the part $\Theta^{\a\bt}_\ve{}\big|_{div}$ of $\Theta^{\a\bt}_\ve$ that diverges as $\ve\ra 0$ in the sense of distributions. Thanks to the assets of our regularization - in particular its manifest reparameterization-  and Lorentz-invariance - this counterterm  entails automatically the properties:
\begin{itemize}
\item[a)] $\Theta^{\a\bt}_\ve{}\big|_{div}$ is covariant, symmetric and traceless;
\item[b)] $\Theta^{\a\bt}_\ve{}\big|_{div}$ is invariant under an arbitrary reparameterization $\la\ra\la'(\la)$ of the world-line;
\item[c)] $\Theta^{\a\bt}_\ve{}\big|_{div}$ is supported on $\G$.
\end{itemize}
The explicit determination of this tensor requires to apply $\Theta^{\a\bt}_\ve$ to a test function $\vp$ of the Schwartz space, to isolate the terms that diverge as $\ve\ra 0$, and to factorize eventually again the test function. The Laurent-expansion of $\Theta^{\a\bt}_\ve(\vp)$ around $\ve=0$ is a bit cumbersome, although conceptually not particularly involved,  see Appendix \ref{dot}. For later convenience we report the resulting expansions of the tensors $A^{\a\bt}$ and $B^{\a\bt}$ separately:
\begin{align}
A^{\a\bt}= \pi\!&\int\! b^2w^2\! \left\{\!-\frac{4b^2}{\ve^4}\,u^\a u^\bt +\frac{1}{\ve^2}
\left(2\eta^{\a\bt}- 4bu^{(\a} \pa^{\bt)}+b^2u^\a u^\bt\, \square\right)\right.\nn\\
&+\left.\ln\ve^2\!\left(\pa^\a\pa^\bt+\frac{\eta^{\a\bt}}{2}\,\square-b u^{(\a} \pa^{\bt)}\square+\frac{b^2}{8}\,u^\a u^\bt\square^2\right)\right\}\dl^4(x-\G(\la,b))\,dbd\la+o(1),\label{aab}\\
B^{\a\bt}=\pi\!&\int\!b^4 \left\{\frac{4w^2}{3\ve^4}\, u^\a u^\bt-\frac{1}{6\ve^2}
\left(w^2u^\a u^\bt\square+2 G^{\a\m}G^{\bt\nu}\pa_\m\pa_\n\right)\right.\nn\\
&+\left.\frac{1}{48}\left(w^2u^\a u^\bt\square + 4 G^{\a\m}G^{\bt\nu}\pa_\m\pa_\n\right)\square \right\}\dl^4(x-\G(\la,b))\,dbd\la+o(\ve).\label{bab}
\end{align}
In these expressions the variables $u$ and  $w$ are evaluated at $\la$, and the
$\dl$-function is supported on the singularity surface $\G^\m(\la,b)$  \eref{Ga}. Correspondingly it is understood that the integration over $\la$ is over the entire real line, while the integration over $b$, we recall, is always restricted to {\it positive} values. The tensor $G^{\a\bt}$ showing up in $B^{\a\bt}$ is defined by
\be\label{gab}
G^{\a\bt}=u^\a w^\bt-u^\bt w^\a,
\ee
and all space-time derivatives $\pa_\m$ are meant applied to the $\dl$-function.

As one sees there are  terms diverging as $1/\ve^4$, $1/\ve^2$ and $\ln \ve$,
but they are all supported on $\G$. The structure of these terms is restricted, apart from Lorentz-covariance, by reparameterization invariance. To check the latter one has to take into account the transformation rules of $b$ and $w^2$  \eref{brep} and \eref{ds}, and to notice that the ``covariant'' tensor \eref{gab} scales as $G'^{\a\bt}=(d\la/d\la')^3 G^{\a\bt}$.

In the expansion of $A^{\a\bt}$ $o(1)$ denotes a distribution that converges as $\ve\ra 0$ in the distributional sense (to a distribution whose support is generically the whole space-time). We will comment on the (innocuous) nature of the logarithmic divergence, not present in  $B^{\a\bt}$, later on.

In the expansion of $B^{\a\bt}$ $o(\ve)$ denotes a distribution that as $\ve\ra 0$ in the distributional sense converges to {\it zero}. For later convenience in the case of $B^{\a\bt}$ we determined also explicitly the finite terms - of order $o(1)$ - although not required for the present purpose that concerns only the divergent terms. The fact that the divergent and finite terms of $B^{\a\bt}$  are all supported on $\G$ is a consequence of  $B^{\a\bt}$ being proportional to $\ve^2$, see \eref{ab}.

Extracting from \eref{aab} and \eref{bab} the divergent terms and inserting them in \eref{the} it is now straightforward to determine the divergent part of $\Theta^{\a\bt}_\ve$, satisfying indeed properties  a)-c),
\begin{align}
\Theta^{\a\bt}_\ve{}\big|_{div}
= \pi\!&\int\!b^2 \left\{\!-\frac{8b^2w^2}{3\ve^4}\,u^\a u^\bt\! +\!\frac{1}{\ve^2}
\left(\!4w^2\eta^{\a\bt}\!- 4bw^2u^{(\a} \pa^{\bt)}\!+\frac{5b^2}{6}\,w^2u^\a u^\bt \square-\frac{b^2}{3}\, G^{\a\m}G^{\bt\nu}\pa_\m\pa_\n
\!\right)\right.\nn\\
&+\left.w^2\ln\ve^2\!\left(\!\pa^\a\pa^\bt+\frac{\eta^{\a\bt}}{2}\,\square-b u^{(\a} \pa^{\bt)}\square+\frac{b^2}{8}\,u^\a u^\bt\square^2\right)\right\}\dl^4(x-\G(\la,b))\,dbd\la.\label{tdiv}
\end{align}
Invoking a ``minimal-subtraction'' scheme we can thus define a (preliminary) renormalized energy-momentum tensor of the electromagnetic field as
\be\label{temi}
t^{\a\bt}_{em} \equiv {\cal S}^\prime \!- \!\lim_{\ve\ra 0}
\left(\Theta^{\a\bt}_\ve -\Theta^{\a\bt}_\ve{}\big|_{div}\right)\!.
\ee
By  construction this distributional limit exists and $t^{\a\bt}_{em}$ satisfies requirements 1)-3). In the next section we will cope with requirement 4), regarding conservation.

\section{Conservation properties and uniqueness}

To explore property 4) we have to determine first of all the (distributional)
four-divergence of the tensor  \eref{temi}.  To do this we take advantage from the fact that derivatives are {\it continuous} operations in ${\cal S}^\prime$. This implies  that we can switch  derivatives with distributional limits so that
\be\label{dtt}
\pa_\a t^{\a\bt}_{em}=  {\cal S}^\prime \!-\!\lim_{\ve\ra 0}
\left(\pa_\a\Theta^{\a\bt}_\ve -\pa_\a\!\left(\Theta^{\a\bt}_\ve{}\big|_{div}\right)\right),
\ee
and it is guaranteed that this limit exists. From \eref{tdiv}, using the operatorial identification $u^\a \pa_\a=-\pa/\pa b$, valid when applied to $\dl^4(x-\G(\la,b))$, one obtains
\be\label{dtdiv}
\pa_\a\!\left(\Theta^{\a\bt}_\ve{}\big|_{div}\right)=
\frac\pi3\!\int\!b^2w^2\! \left(-\frac{32b}{\ve^4}\,u^\bt  +\frac{2}{\ve^2}\,\big(2bu^\bt\square-3\pa^\bt\big)\!\right)\dl^4(x-\G(\la,b))\,dbd\la.
\ee
The four-divergence of $\Theta^{\a\bt}_\ve$ can be computed from \eref{the} and \eref{ab} - using $L^2=\ve^2$ and $\pa_\a \la_\ve(x)= L_\a/(uL)$ - and reads
\be\label{divreg}
\pa_\a\Theta^{\a\bt}_\ve=\ve^2 \left(\frac{2w^2(wL)}{(uL)^6 } - \frac{(wB)}{(uL)^5}\right)L^\bt, \quad\quad B^\m\equiv \frac {dw^\m}{d\la}.
\ee
Notice that - in agreement with \eref{pw} - this four-divergence is proportional to $\ve^2$. In fact, since on general grounds in the complement of $\G$ the {\it naive} energy-momentum \eref{t0} satisfies the ``free'' conservation law $\pa_\a\Theta^{\a\bt}=0$, in the complement of $\G$ $\pa_\a\Theta^{\a\bt}_\ve$ must converge {\it point-wise} to zero.

To proceed we must expand the r.h.s. of \eref{divreg} in powers of $\ve$ (see Appendix \ref{fdo})
\be\label{d4e}
 \pa_\a\Theta^{\a\bt}_\ve=
\frac \pi3\!\int\! b^2w^2\!\left(\!-\frac{32b}{\ve^4}\,u^\bt +\frac{2}{\ve^2}\,
\big(2bu^\bt\square -3\pa^\bt\big) -\frac12\,\big(bu^\bt\square -3\pa^\bt\big)\square\right)\!\dl^4(x-\G(\la,b))\,dbd\la +o(\ve).
 \ee
Inserting these results in \eref{dtt} one sees that the divergent terms cancel - as they must do by construction - and the result is
\be\label{divt}
\pa_\a t^{\a\bt}_{em}=
 \frac{\pi}{6}\!\int\!b^2w^2\big(3\pa^\bt -bu^\bt\square\big)\square\,\dl^4(x-\G(\la,b))\,dbd\la.
\ee

\subsection{A non-minimal subtraction}

So far we have constructed a renormalized energy-momentum tensor that satisfies requirements 1)-3), but not 4), {\it i.e.} its four-divergence is not supported on $\g$, but rather on $\G$. To cure this problem we resort to the indeterminacy related to the addition of finite counterterms ${ D}^{\a\bt}$ supported on $\G$. These terms must be chosen such that the modified energy-momentum tensor satisfies condition 4) and, actually, it is not difficult to find one. Introducing the (traceless and reparameterization invariant) tensor supported on $\G$
\be\label{dab}
{ D}^{\a\bt}_{(0)}\equiv \pi\!\int\!b^2\!\left(\!-w^2\pa^\a\pa^\bt + \frac{w^2}{2}\,\eta^{\a\bt}\square+\frac{b^2w^2}{24}\,u^\a u^\bt\square^2-\frac{b^2}{12}\, G^{\a\m}G^{\bt\nu}\pa_\m\pa_\n\right)\!\dl^4(x-\G(\la,b))\,dbd\la,
\ee
a simple calculation gives in fact
\be \label{ddab1}
\pa_\a { D}^{\a\bt}_{(0)}= -\frac{\pi}{6}\!\int\!b^2w^2\big(3\pa^\bt -bu^\bt\square\big)\square\,\dl^4(x-\G(\la,b))\,dbd\la,
\ee
which is precisely the opposite of \eref{divt}. Enforcing  a {\it non-minimal} subtraction, from \eref{temi}, \eref{divt} and \eref{ddab1} it follows  that the renormalized energy-momentun tensor of the electromagnetic field generating by a massless charged particle - satisfying  properties 1)-4) - is given by
\begin{align}
T^{\a\bt}_{em}  \equiv t^{\a\bt}_{em} +{ D}^{\a\bt}_{(0)}& = {\cal S}^\prime \!- \!\lim_{\ve\ra 0}
\left(\Theta^{\a\bt}_\ve -\Theta^{\a\bt}_\ve{}\big|_{div}+{ D}^{\a\bt}_{(0)}\right)
\label{line1}\\
&
={\cal S}^\prime \!-\!\lim_{\ve\ra 0}\left(A^{\a\bt}+ H^{\a\bt}\right)\!,\label{temh}
\end{align}
where $A^{\a\bt}$ is the {\it bare} energy-momentum tensor introduced in \eref{ab},  and the {\it total} counterterm $H^{\a\bt}$ has the expression
\begin{align}
H^{\a\bt}&=\pi\!
\int\!b^2w^2\!\left\{\!\frac{4b^2}{\ve^4}\,u^\a u^\bt\! -\!\frac{1}{\ve^2}
\left(2\eta^{\a\bt}- 4bu^{(\a} \pa^{\bt)}+b^2u^\a u^\bt \square\right)
-\ln\ve^2\!\left(\!\pa^\a\pa^\bt+\frac{\eta^{\a\bt}}{2}\,\square\right.\right.\nn\\
&\phantom{........}\left.\left.-b u^{(\a} \pa^{\bt)}\square+\!\frac{b^2}{8}\,u^\a u^\bt\square^2\!\right)\!
+\frac{b^2}{16}\,u^\a u^\bt\square^2\!-\pa^\a\pa^\bt
\right\}\dl^4(x-\G(\la,b))\,dbd\la.\label{line3}
\end{align}
To obtain \eref{temh}, in  \eref{line1}  we used expressions \eref{the}-\eref{L2e},
\eref{tdiv} and \eref{dab} and inserted for $B^{\a\bt}$ the expansion \eref{bab}.
From \eref{divt}, \eref{ddab1} and \eref{line1} we deduce that  $T^{\a\bt}_{em}$ satisfies the basic identity
\be\label{dtmn}
 \pa_\a T^{\a\bt}_{em}=0,
\ee
meaning that \eref{dtloc} holds with vanishing self-force, {\it i.e.} $f^\bt=0$. Correspondingly equation \eref{eqm} furnishes as equation of motion of a ``self-interacting'' massless  charge  - in absence of external fields - the equation of {\it free} motion
\be
\frac{dp^\m}{d\la}=0,
\ee
to be compared with \eref{LD} for a massive particle.

\subsection{Uniqueness and finite counterterms}

As in the massive case  there remains the possibility to modify the tensor $T^{\a\bt}_{em}$ \eref{line1} by further finite counterterms ${ D}^{\a\bt}$ subject to the  constraints:
\begin{itemize}
\item[a)] ${ D}^{\a\bt}$ is a distribution supported on $\G$, with length dimension  $1/L^4$;
\item[b)] ${ D}^{\a\bt}$ is  covariant, symmetric,  traceless and reparameterization invariant;
\item[c)] the four-divergence of ${ D}^{\a\bt}$ is multiplicatively supported on $\g$, {\it i.e.}
\be\label{dcal}
  \pa_\a{ D}^{\a\bt}  =\int \!{\cal G}^\bt(\la)\, \dl^4(x-y(\la))\,d\la,
    \ee
for some vector ${\cal G}^\bt$.
\end{itemize}
The general form of these counterterms is thus
\be\label{dd}
{ D}^{\a\bt}=\int\! d^{\a\bt}\,\dl^4(x-\G(\la,b))\,dbd\la,
\ee
where the tensors $d^{\a\bt}$ depend on $b$ and $\la$ and may involve also derivative operators acting on the $\dl$-function; see below for explicit examples.

The most efficient way to search for tensors satisfying a)-c) consists in searching  first for covariant vectors ${\cal G}^\bt\equiv {\cal G}^\bt(\la)$. Due to a) and b) these vectors must have length dimension $1/L^2$, and under a reparameterization \eref{repara} they must transform as ${\cal G}'^\bt=(d\la/d\la')\,{\cal G}^\bt$. Relying again on the {\it regularity paradigm} introduced in Section \ref{uniq1}, that forbids powers of the acceleration $w^\m$ or its derivatives at the denominator, ${\cal G}^\bt$ can then only be of the  (operatorial) form
\[
{\cal G}^\bt = c\, u^\bt \square,
\]
with $c$ a constant. But then  $\pa_\a{ D}^{\a\bt}$ is not {\it multiplicatively} supported on $\g$, unless  $c=0$. This means that $ {\cal G}^\bt$  must vanish so that property c) simplifies to
\be\label{ddab}
  \pa_\a{ D}^{\a\bt}=0.
\ee
The counterterms  ${ D}^{\a\bt}$ induce therefore {\it physically irrelevant} modifications to
$T^{\a\bt}_{em}$, in that  thanks to \eref{ddab} the modified tensor $T^{\a\bt}_{em} + { D}^{\a\bt}$ leads to the same equation of motion of the particle as  $T^{\a\bt}_{em}$ (see Section \ref{eom}). In conclusion, the energy-momentum tensor \eref{line1} is unique,  modulo physically irrelevant finite counterterms \cite{unb}.

Despite their conceptually limited relevance one may ask if there exist non-vanishing tensors $D^{\a\bt}$ of the form \eref{dd}, satisfying conditions  a) and b) and equation \eref{ddab}. The answer is affirmative  and one may try a classification.
The corresponding tensors $d^{\a\bt}$ must have length dimension $1/L^2$ and under a reparameterization they must have {\it weight} $-2$, see \eref{brep},
\[
d'^{\a\bt}=\bigg(\frac{d\la}{d\la'}\bigg)^2 d^{\a\bt}.
\]
Reparameterization-covariant objects are for example, see \eref{gab},
\be\label{obj}
b, \quad \quad bu^\a,\quad\quad  u^\a,\quad\quad G^{\m\n},  \quad\quad w^2, \quad\quad \pa^\m,
\ee
with weight respectively $1$, $0$, $-1$, $-3$, $-4$ and $0$. Notice however that $w^\m$ itself does not transform covariantly under reparameterizations.

The  tensors $d^{\a\bt}$ for which \eref{dd} satisfies conditions a) and b) and \eref{ddab} are classified by the powers of $w$ and/or its successive derivatives. There are no such $d^{\a\bt}$ of zero order in $w$. At first order in $w$ we find the unique solution
\be\label{d0}
D_1^{\a\bt}=  \int\! b\left(bu^{(\a} G^{\bt)\m}\square+2\, G^{\m(\a}\pa^{\bt)}\right)\!\pa_\m
\dl^4(x-\G(\la,b))\,dbd\la,
\ee
and  at second order in $w$ we find the two independent solutions
\begin{align}
D_2^{\a\bt}&=  \int\! b^2w^2 \!\left(\!\pa^\a\pa^\bt+\frac{\eta^{\a\bt}}{2}\,\square-b u^{(\a} \pa^{\bt)}\square+\frac{b^2}{8}\,u^\a u^\bt\square^2\right)  \dl^4(x-\G(\la,b))\,dbd\la,\label{d1}
\\
D_3^{\a\bt}&=  \int\! b^2\!\left(b^2 G^{\a\m}G^{\bt\nu}\pa_\m\pa_\n\square+4w^2(\pa^\a\pa^\bt-\eta^{\a\bt}\square)\right)
\dl^4(x-\G(\la,b))\,dbd\la.\label{d2}
\end{align}
To check that these tensors are traceless and divergenceless one must use the operatorial identification $u^\a \pa_\a\equiv -\pa/\pa b$, valid when applied to $\dl^4(x-\G(\la,b))$.
Notice that $D_2^{\a\bt}$ corresponds precisely to the logarithmic divergence in \eref{tdiv}; this is the reason for why the logarithmic divergence cancels out from \eref{dtdiv}.

It is not difficult to realize that for each of these counterterms there exists a ``local'' tensor  \be\label{kab0}
K^{\g\a\bt}=\int\! k^{\g\a\bt}\,\dl^4(x-\G(\la,b))\,dbd\la,\quad\quad k^{\g\a\bt}=- k^{\a\g\bt},
\ee
such that
\be\label{kab}
D^{\a\bt}=\pa_\g K^{\g\a\bt},
\ee
that trivializes hence the property \eref{ddab}. One has for example $D^{\a\bt}_2=\pa_\g K^{\g\a\bt}_2$,
with
$$
K_2^{\g\a\bt}= \int\! b^2w^2 \!\left(\!-\eta^{\bt[\g} \,\pa^{\a]}+b\, u^{[\g}\pa^{\a]}\pa^\bt-\frac{b^2}{4}\,u^\bt u^{[\g}\pa^{\a]}\,\square\right)\dl^4(x-\G(\la,b))\,dbd\la.
$$

Since \eref{ddab} is always satisfied as an  ``algebraic'' identity, we conjecture that all finite counterterms  are of the form \eref{kab}, \eref{kab0}, and correspond thus to a ``classical'' indeterminacy of energy-momentum tensors in field theory.

 \subsection{Renormalized energy-momentum tensor for unbounded trajectories}\label{ut}

With respect to a bounded trajectory the field \eref{unbound} of an unbounded trajectory acquires an additional term: the shock-wave  $F^{\m\n}_{sw}$   in \eref{fsw}, that is proportional to
$\dl(v_\infty x)$. This field is thus non-vanishing only on a plane  traveling at the speed of light, orthogonal to the trajectory  of a virtual particle in linear motion
with world-line ${\cal L}$, parameterized by $y^\m_{\cal L}(\la)=v^\m_\infty\, \la=(1,\vec v_\infty)\la$.

We consider first the simplest such case, {\it i.e.} a strictly linear motion - $w^\m(\la)=0$ for all $\la$ - for which the field is a pure shock-wave: $F^{\m\n}=F^{\m\n}_{sw}$. Since this field is almost everywhere vanishing and the four-momentum of the particle is conserved, the energy-momentum tensor $T^{\a\bt}_{em}$ must be a distribution that is $i)$ covariant, symmetric and traceless, $ii)$ proportional to $\dl(v_\infty x)$ and $iii)$ divergenceless. Since it must be constructed  with $x^\m$ and $v^\m_\infty$, it is immediately seen that no such distribution exists.
In conclusion, for a pure shock-wave $T^{\a\bt}_{em}$ must vanish.

For a generic unbounded motion the singularity-locus of the field $F^{\m\n}$ \eref{unbound} - w.r.t. the bounded case - is enriched by the shock-wave-plane, and the singularity curve $\vec\G$ in \eref{ga} starts now from the particle's trajectory $\g$ and ends on ${\cal L}$. Since the tensor $T^{\a\bt}_{em}$ in \eref{line1} is a (divergenceless) distribution also for unbounded motions, according to our general procedure 1)-4) - specifically condition 3) -  with respect to this tensor there could now appear new additional {\it finite} counterterms supported on the shock-wave-plane or on the intersection of $\G$ and the shock-wave-plane, {\it i.e.} the line ${\cal L}$. However, since the four-divergence of these counterterms must be supported on $\g$, and neither the  shock-wave-plane nor ${\cal L}$ intersect $\g$, their four-divergence must be zero. They must then vanish for the same reasons as for the linear motion considered above.

In conclusion the renormalized energy-momentum tensor of the electromagnetic field produced by an unbounded motion is  still given by \eref{line1} and satisfies, in particular, the conservation  law \eref{dtmn}. The main difference w.r.t. to the bounded case is that - the singularity string having a finite extension - the counterterms \eref{tdiv}, \eref{dab} and \eref{line3} are all of compact spatial support. Moreover, since the acceleration of the particle is supposed to vanish sufficiently fast at past infinity, see \eref{asym}, the tensor \eref{line1} has a falloff at spatial infinity that makes the total four-momentum integrals convergent (see Section \ref{mi}).

\subsection{Interpretation: massless charges do not radiate}

Considering the tensor  $T^{\a\bt}_{em}$ in the form \eref{temh} we observe first of all that the {\it naive}  expression $A^{\a\bt}= -w^2L^\a L^\bt/(uL)^4$
- that is frequently used in the literature to analyze radiation reaction for massless charges - as it stands is meaningless: it does neither represent a distribution, nor does it have well-defined conservation properties. First one must regularize it in some way - we have chosen  \eref{rte} - and before taking the limit of $\ve\ra 0$ one must add the counterterm  $H^{\a\bt}$ supported on $\G$,  encoding divergent as well as finite contributions. Only in the complement of $\G$ $T^{\a\bt}_{em}$ coincides with $A^{\a\bt}$. While the divergent contributions of  $H^{\a\bt}$ - of the type  $1/\ve^4$, $1/\ve^2$ and $\ln \ve$ - are needed to ensure the existence of the distributional limit \eref{temh},  the finite ones are necessary to ensure property 4) - related with covariance and total four-momentum conservation:  the preliminary energy-momentum tensor   $t^{\a\bt}_{em}$ \eref{temi} - satisfying \eref{divt} - would in fact give rise to (finite) four-momentum integrals $\int\! t^{0\bt}_{em}\,d^3x$ that are {\it not} four-vectors - a statement that we will cross-check explicitly in Section \ref{mi}.

The basic result \eref{dtmn} states that the four-momentum of the electromagnetic field is conserved independently from the one of the particle - a conclusion with a far reaching consequence: a classical massless charged particle does not ``communicate'' four-momentum to its electromagnetic field, {\it i.e.} it does not
{\it not} emit {\it bremsstrahlung}, or more generically radiation. We will comment this basic conclusion further in the forthcoming  sections.

\section{Four-momentum integrals for unbounded trajectories}\label{mi}

From the mathematical point of view a further virtue of our distributional approach is that the basic limits \eref{line1}, \eref{temh} exist also in a ``stronger'' sense than in ${\cal S}^\prime(\mathbb{R}^4)$, {\it i.e.} they exist also at fixed time in the topology of ${\cal S}^\prime(\mathbb{R}^3)$.  This is due to the fact that the singularities of the electromagnetic field \eref{unbound} are all ``space-like'', {\it i.e.} they occur at fixed time in a specific spatial region.

This means that the above limits hold also if we apply them at fixed time $t$ to a test function $\vp(\vec x) \in{\cal S}(\mathbb{R}^3)$, and that in the relations involving derivatives, such as \eref{dtmn}, the derivative w.r.t. time can be treated as a parametric derivative. Moreover, if we consider unbounded trajectories with (sufficiently fast) vanishing accelerations in the infinite past,
\be\label{ww}
\lim_{\la\ra-\infty} w^\m(\la)=0,
\ee
as anticipated at the end of Section \ref{ut} the tensors  $A^{\a\bt}$ and $H^{\a\bt}$ in \eref{temh} admit finite integrals over whole space, so that we can  enlarge the space of test functions ${\cal S}(\mathbb{R}^3)$ to include functions that at infinity become constant, in particular the constant function $\vp_0(\vec x)=1$. This means that for unbounded motions the total four-momentum of the field
\be\label{pem}
P^\bt_{em}(t)= T^{0\bt}_{em}(\vp_0)= \int\! T^{0\bt}_{em}(t,\vec x)\,d^3x
\ee
is {\it finite} for every $t$. Applying likewise  equation \eref{dtmn} to  $\vp_0(\vec x)$
we obtain
\[
\frac{d}{dt}\int\! T^{0\bt}_{em}(t,\vec x)\,d^3x +\int\! \pa_i T^{i\bt}_{em}(t,\vec x)\,d^3x =0,
\]
that through Gauss's law leads to
\[
\frac{dP^\bt_{em}(t)}{dt}=-\int\! T^{i\bt}_{em}(t,\vec x)\,d\Sigma^i=0.
\]
The last term - the flux of $T^{i\bt}_{em}$ across a sphere with radius $R$ tending to infinity - vanishes because $i)$ the tensor $A^{\a\bt}$  in \eref{temh}, being multiplied by $w^2$, vanishes rapidly at infinity \cite{unb1} and $ii)$ the tensor $H^{\a\bt}$ for every $t$ is of compact spatial support.

We conclude thus that the total four-momentum  $P^\bt_{em}(t)$  of the field is a constant. To determine it we can thus evaluate it in the limit $t\ra-\infty$. Inserting \eref{temh} in \eref{pem}, as $t\ra-\infty$ thanks to \eref{ww} both terms of \eref{temh} give rise to integrals that converge to zero and we conclude thus that, actually,
\be\label{pt}
P^\bt_{em}(t)=0.
\ee

\subsection{Explicit evaluation of the total four-momentum}

In this section we cross-check the prediction \eref{pt} via an explicit computation. In doing  this we exemplify also how the - apparently abstract - definition \eref{temh} is actually {\it operative}, in that it allows to determine concretely  the four-momentum $P^\bt_{em,V}(t)=\int_V T^{0\bt}_{em}(t,\vec x)\,d^3x$ contained at time $t$ in a generic volume $V$ - whether $V$ intersects/contains the singularity line \eref{ga} or not. This analysis is also instructive because it reveals what we would have obtained for the total four-momentum, would we not have added the counterterm $H^{\a\bt}$.

According to  \eref{pem} and \eref{temh}  the total four-momentum at fixed time $t$ is given  by
\be\label{limp}
P^\bt_{em}(t)=\lim_{\ve\ra 0}\left(\int \!A^{0\bt}(t,\vec x)\,d^3x+ \int \!H^{0\bt}(t,\vec x)\,d^3x\right) \equiv \lim_{\ve\ra 0}\left(P^\bt_A(t)+P^\bt_H (t) \right).
\ee
We begin evaluating the four-momentum of the bare energy-momentum tensor. As shown in appendix \ref{fmi}, after integrating over angles one ends up with the one-dimensional integral
\be\label{pat}
P^\bt_A(t)=\frac{4\pi}{\ve^4}\int_0^\infty\! a^2(\tau_\ve)\! \left(r^2+\ve^2,r^2\,\vec v(\tau_\ve)\right)r^2dr,
\ee
where  $\vec v$ and $\vec a$ are evaluated at the ``retarded'' time $\tau_\ve=t-\sqrt{r^2+\ve^2}$. The integral is convergent in that for $r\ra \infty$ we have $a(\tau_\ve)\approx a(-r)$ and -  thanks to \eref{ww} -  for large negative values  the acceleration vanishes rapidly. Expression \eref{pat} has a clear physical meaning:
it is the {\it naiv} (diverging) total four-momentum of the electromagnetic field, {\it i.e.} before renormalization.

Since eventually we have to take the limit $\ve\ra 0$ we expand \eref{pat} in inverse powers of $\ve$:
\be\label{patt}
P^\bt_A(t)=
\frac{4\pi}{\ve^4}\!\int_0^\infty\!\! a^2(\tau)(1,\vec v(\tau))\,r^4dr-
\frac{2\pi}{\ve^2}\!\int_0^\infty\!\! a^2(\tau)(1,3\vec v(\tau))\,r^2dr-
\frac{\pi}{2}\!\int_0^\infty\! \!a^2(\tau)(1,-3\vec v(\tau))\,dr+o(\ve),
\ee
where $\tau=t-r$. We have thus two contributions diverging respectively as $1/\ve^4$ and $1/\ve^2$, and a finite one. Notice that although the leading divergence of \eref{patt} can be cast in the, at first glance, covariant form (parameterizing the world-line with time, so that $w^2=-a^2$)
\be\label{paa}
-\frac{4\pi}{\ve^4}\!\int_0^\infty\! w^2(\tau)\,u^\bt(\tau)\,r^4dr,
\ee
due to the special role played  by the time-coordinate it is {\it not} covariant at
all \cite{lead}. Actually $P^\bt_A(t)$ is neither a four-vector, nor is it conserved.

To evaluate the four-momentum coming from the counterterm we rewrite  the $0\bt$-components of \eref{line3} choosing as parameter $\la=y^0(\la)$ and integrate then out the temporal $\dl$-function:
\begin{align}
H^{0\bt}(t,\vec x)=
& -\frac{4\pi}{\ve^4}\int \!b^4a^2\, u^\bt\,\dl^3\big(\vec x-\vec\G(t,b)\big)db
\nn\\
  &+ \frac{\pi}{\ve^2}\left(\!2 \eta^{0\bt}\!\!\int\! b^2a^2\dl^3\big(\vec x-\vec\G(t,b)\big)db -2\pa^\bt \!\int\! b^3a^2 \dl^3\big(\vec x-\vec\G(t,b)\big)db\right.\nn\\
&\left.\phantom{........}-2\pa^0\!\!\int\! b^3a^2 u^\bt \dl^3\big(\vec x-\vec\G(t,b)\big)db+\square\!\int\! b^4a^2 u^\bt \dl^3\big(\vec x-\vec\G(t,b)\big)db\right)\nn
\\
&+\pi\ln\ve^2\!\left(\pa^0\pa^\bt \!\int \!b^2a^2\dl^3\big(\vec x-\vec\G(t,b)\big)db +\frac{\eta^{0\bt}}2\,\square\!\int\! b^2a^2 \dl^3\big(\vec x-\vec\G(t,b)\big)db\right.\nn\\
&\left.\phantom{...........}-\frac12\,\pa^\bt\square \!\int\! b^3a^2 \dl^3\big(\vec x-\vec\G(t,b)\big)db-\frac12\,\pa^0\,\square\!\int\! b^3a^2 u^\bt \dl^3\big(\vec x-\vec\G(t,b)\big)db\right.
\nn\\
&\left.\phantom{...........}+\frac18\,\square^2\! \int \!b^4a^2 u^\bt\dl^3\big(\vec x-\vec\G(t,b)\big)db
\right)\nn\\
&
-\frac{\pi}{16}\,\,\square^2\! \int \!b^4a^2 u^\bt\dl^3\big(\vec x-\vec\G(t,b)\big)db
+\pi\,\pa^0\pa^\bt\!\int \!b^2a^2 u^\bt\dl^3\big(\vec x-\vec\G(t,b)\big)db.
\nn
\end{align}
In this expression - as in \eref{tdiv} -  it is again understood that $b$ is integrated over the positive real axis, and $\vec\G(t,b)$ is the singularity curve \eref{ga}. The vector  $u^\bt$ stands for $(1,\vec v)$ and the variables $\vec a$ and $\vec v$ are evaluated at time $t-b$. When integrating this expression over whole space the $\dl^3$-functions integrate to unity, but whenever there is a spatial derivative $\partial_i$ in front of the integrals - thanks to Gauss's law and the fact that  $\dl^3\big(\vec x-\vec\G(t,b)\big)$ at fixed $t$ is of compact spatial support - the result is zero. On the other hand the temporal derivatives $\pa^0=\pa/\pa t$ - once one has integrated over space - act on $\vec a(t-b)$ and $\vec v(t-b)$  and they turn therefore into the derivatives $-\pa/\pa b$, that eventually can be integrated by parts. In particular, since  the terms multiplying $\ln\ve^2$ correspond to the  divergenceless counterterm \eref{d1}, they must cancel out when integrated over whole space, as can be checked explicitly. The final result for $P^\bt_H (t)= \int \!H^{0\bt}(t,\vec x)\,d^3x$ reads
\[
P^\bt_H(t)=
-\frac{4\pi}{\ve^4}\!\int_0^\infty\!\!b^4 a^2(\tau)\big(1,\vec v(\tau)\big)\,db+
\frac{2\pi}{\ve^2}\!\int_0^\infty\!\!b^2 a^2(\tau)\big(1,3\vec v(\tau)\big)\,db+
\frac{\pi}{2}\!\int_0^\infty\! \!a^2(\tau)\big(1,-3\vec v(\tau)\big)\,db,
\]
where $\tau=t-b$.
As we see,  $P^\bt_A (t)= -P^\bt_H (t)+o(\ve)$, so that the limit \eref{limp} gives
$P^\bt_{em} (t)=0$, as foreseen in \eref{pt}.

\section{Equations of motion and absence of radiation reaction}\label{eom}

\subsection{External field and vanishing self-force}\label{efa}

To implement our strategy \eref{dtloc}-\eref{eqm} to the derive the equation of motion of the particle in a non-trivial case, we introduce an external (regular) $C^\infty$-field ${\cal F}^{\m\n}$ satisfying the homogeneous equations \eref{maxf}. To keep the total four-momentum finite we consider again an unbounded motion, and correspondingly we choose an external field of compact spatial support.

In this case the  formal electromagnetic energy-momentum tensor is
\be\label{text}
\Theta^{\a\bt}=(F+{\cal F}|F+{\cal F})^{\a\bt},
\ee
where  $F^{\m\n}$ is the field \eref{unbound} of an unbounded motion.
Since the external field is regular and $F^{\m\n}$ is a distribution, also the mixed term $2(F|{\cal F})^{\a\bt}$ is a distribution, as is of course also  $({\cal F}|{\cal F})^{\a\bt}$. Accordingly in presence of an external field the electromagnetic energy-momentum tensor satisfying properties 1)-4) is
\be\label{temex1}
\mathbb{T}^{\a\bt}_{em}= T^{\a\bt}_{em}
+2\,(F|{\cal F})^{\a\bt}+ ({\cal F}|{\cal F})^{\a\bt},
\ee
where ${T}^{\a\bt}_{em}$ is given in \eref{temh}.
Off the singularity-locus we have again $\mathbb{T}^{\a\bt}_{em}= \Theta^{\a\bt} $. Notice, however, that the mixed term $2(F|{\cal F})^{\a\bt}$ contains $\dl$-like terms supported on $\G$, as well as  $\dl$-like terms supported on the shock-wave (see the fields $F^{\m\n}_{sing}$ \eref{bound} and  $F^{\m\n}_{sw}$ \eref{fsw}). These terms are actually essential to cope with requirement 4), {\it i.e.} the relation \eref{dtloc}.
Using the Leibnitz-rule \eref{rule}, from \eref{temex1}, \eref{dtmn} and the fact that the field \eref{unbound} satisfies the equation $\pa_\m F^{\m\n}=j^\n$, we find indeed
\be\label{dtemex}
\pa_\a \mathbb{T}^{\a\bt}_{em}=-
e \!\int\!{\cal F}^{\bt\n}u_\n\, \delta^4(x-y(\la))\,d\la.
\ee
Without the terms of \eref{temex1} supported on $\G$ and on the shock-wave the four-divergence $\pa_\a \mathbb{T}^{\a\bt}_{em}$ would, in fact, not be supported on $\g$.
For the total energy-momentum tensor
$$
{T}^{\a\bt}= \mathbb{T}^{\a\bt}_{em}+  \int \!u^\a p^\bt \dl^4(x-y(\la))\,d\la
$$ equation \eref{dtemex} gives
\be
\pa_\a T^{\a\bt}= \int\!\left(\frac{dp^\bt}{d\la}-e{\cal F}^{\bt\n}u_\n\!\right)\dl^4(x-y(\la))\,d\la.
\ee
Local four-momentum conservation implies thus that the particle must fulfill the {\it bare} Lorentz-equation (to compare with the Lorentz-Dirac equation \eref{ldf} for a  massive particle)
\be\label{dpext}
\frac{dp^\m}{d\la}=e{\cal F}^{\m\n}u_\n.
\ee
This leads to the - {\it a priori} unexpected - conclusion that a particle also in presence of an external field {\it experiences no radiation reaction}. This is of course in line with the fact - expressed by the identities  \eref{dtmn} and \eref{pt} - that the field created by the particle carries vanishing total four-momentum.

Proceeding as in Section \ref{tfm} from the equations above we get the formula for the total conserved four-momentum
\be\label{ggg}
 P^\bt=\int
{T}^{0\bt}(t,\vec x)\,d^3x=  p^\bt(t)-e
 \int_{-\infty}^{\la(t)}{\cal F}^{\bt\n}u_\n\,d\la+P^\bt_{ext},
\ee
where the four-momentum $P^\bt_{ext}=\int ({\cal F}|{\cal F})^{0\bt}d^3x$ of the external field is again separately conserved, to be compared with the corresponding expression \eref{ptotm} for the massive case.

Equation \eref{dpext} is  in net contradiction with equation \eref{post} - derived in Section \ref{self} at the basis of an ``arbitrary'' regularization/renormalization prescription for the divergent self-interaction. This prescription had however - we recall -  no {\it a priori} fundamental motivation and its character is nothing more than {\it heuristic}. A part from this, the putative self-force \eref{f0} contains inverse powers of the acceleration - the largest one being  $1/(-w^2)^{13/4}$ - meaning that this force does not admit a ``flat'' limit, {\it i.e.} it diverges whenever a particle follows a linear motion - a highly non-physical behavior. On the contrary equation \eref{dpext} is {\it regular} and foresees a {\it vanishing} self-force.

\subsection{Interacting particles}

A system playing a central role in Electrodynamics is that of an isolated set of interacting charged particles. As a prototypical case we consider two massless particles - a particle 1 with charge $e_1$, world-line $y_1^\m(\la)$ etc., and a particle 2 with charge $e_2$, world-line $y_2^\m(\la)$ etc. The formal electromagnetic energy-momentum tensor of this system is
\be\label{t12}
\Theta^{\a\bt}=(F_1+F_2|F_1+F_2)^{\a\bt},
\ee
where the fields $F_1^{\m\n}$ and  $F_2^{\m\n}$  have the form \eref{unbound}. We suppose that both particles follow {\it unbounded} trajectories, approaching in the infinite past sufficiently fast linear motions.

According to our strategy 1)-4), to construct a renormalized energy-momentum tensor we must first identify the singularity-locus. With this respect we recall an assumption that is usually {\it implicitly} made in the electrodynamics of massive particles:
it is assumed that the trajectories of the particles never intersect. This hypothesis originates from the fact that at a particle's position the field is infinite, and it is justified because the ``probability'' of such an intersection is zero, meaning that it happens only for ``exceptional'' motions.

For massless particles we impose a similar {\it Dirac-veto}: a particle - massless or not - can never hit the singularity string of another massless particle. The reason is, of course, that at those positions the electromagnetic  fields diverge and the justification arises again from the fact that the probability of such intersections is zero. Notice  that this veto implies, in particular, that the particles' trajectories themselves never intersect.

On the contrary, the probability that, say, particle 1 hits the shock-wave plane of particle 2 - with equation $(v_{2\infty} x)=v_{2\infty}^{\m} x_\m=0$, see \eref{fsw} -
is of order {\it unity}, in that generically in $D=4$ a curve (the world-line of  particle 1) {\it does} intersect a tree-dimensional manifold  (the three-dimensional hypersurface swept out by the shockwave produced by particle 2).
However, as long as these intersections are {\it generic}, the products of  $F_1^{\m\n}$ and  $F_{2}^{\m\n}$ are distributions and, moreover, the derivatives of these products can be computed using the Leibnitz-rule.

In conclusion the renormalized energy-momentum tensor of the electromagnetic field  is given by
\be\label{tem12}
\mathbb{T}^{\a\bt}_{em}=T^{\a\bt}_{(1)em} +T^{\a\bt}_{(2)em}+2(F_1|F_2),
\ee
where the terms  $T^{\a\bt}_{(i)em}$ are those in \eref{temh}. Using \eref{dtmn} for both particles, and applying the Leibnitz-rule \eref{rule} together with the respective Maxwell equations satisfied by $F_1^{\m\n}$ and $F_2^{\m\n}$, from  \eref{tem12} we derive the identity (in the tensors $T^{\a\bt}_{(i)em}$ we must restore a factor $(e_i/4\pi)^2$)
\be\label{d12}
 \pa_\a \mathbb{T}^{\a\bt}_{em}=
-e_1\!\int\!{ F}^{\bt\n}_{2}(y_1)\,u_{1\n}\, \delta^4(x-y_1)\,d\la
- e_2\!\int\!{ F}^{\bt\n}_{1}(y_2)\,u_{2\n} \,\delta^4(x-y_2)\,d\la.
\ee
Imposing that the total energy-momentum tensor
\be\label{ttot1}
{T}^{\a\bt}= \mathbb{T}^{\a\bt}_{em} + \int \!u^\a_1 p^\bt_1 \dl^4(x-y_1)\,d\la+
\int \!u^\a_2 p^\bt_2 \dl^4(x-y_2)\,d\la
\ee
is conserved,  $\pa_\a{T}^{\a\bt}=0$, eventually we obtain the {\it bare}
Lorenz-equations of motion
\begin{align}
\frac{dp^\m_1}{d\la}&=e_1{ F}^{\m\n}_2(y_1)u_{1\n},\label{p1}\\
\frac{dp^\m_2}{d\la}&= e_2{ F}^{\m\n}_1(y_2)u_{2\n}.\label{p2}
\end{align}

\subsubsection{Total four-momentum}

Proceeding as in Section \ref{tfm}, from the formulae above, in particular \eref{d12}, we can again derive an explicit expression for the conserved total four-momentum $P^\bt$ of the two-particle system.

As seen in Section \ref{mi}, the first two terms in \eref{tem12} give a vanishing contribution to $P^\bt$. The contribution of the (third) mixed term of \eref{tem12} can be read off from \eref{d12}, as in Section \ref{tfm}. We obtain thus the - formally natural - result
\be
 P^\bt=p_1^\bt(t)+p_2^\bt(t) -e_1
 \int_{-\infty}^{\la_1(t)}\!F_2^{\bt\n}(y_1)u_{1\n}\,d\la
  -e_2
 \int_{-\infty}^{\la_2(t)}\!F_1^{\bt\n}(y_2)u_{2\n}\,d\la.\label{mix12}
\ee
A peculiarity arises, however, from the interaction terms in \eref{mix12}, in that each field $F_i^{\m\n}$ is composed of the three fields \eref{unbound}. To be specific we consider the last term in \eref{mix12}, involving  $F_1^{\m\n}$.
The regular part of this field, $F^{\m\n}_{1reg}$, gives rise to a {\it continuous} contribution to  $P^\bt$. The $\dl$-like term $F^{\m\n}_{1sing}$ drops out from \eref{mix12}, since - thanks to the Dirac-veto - particle 2 never hits the singularity string of particle 1. The shock-wave field $F^{\m\n}_{1sw}$  \eref{fsw} gives instead rise to the {\it discontinuous} contribution to \eref{mix12}
  \be \label{mix}
-e_2\int_{-\infty}^{\la_2(t)}\!F_{1sw}^{\bt\n}(y_2)\,u_{2\n}\,d\la= -\frac{e_1e_2}{2\pi} \,
\sum\,\frac {(y_2u_2)v_{1\infty}^\bt-(v_{1\infty} u_2) y_2^\bt}
 {y_2^2\,|(v_{1\infty} u_2)|},
\ee
where the sum is over all intersections of particle 2 with the shock-wave plane -  $(v_{1\infty} y_2)=t-\vec v_{1\infty}\cdot\vec y_2(t)=0$ - occurring {\it before} time $t$.  This means that the last term in \eref{mix12} jumps discontinuously - by a finite amount - whenever particle 2 crosses the shock-wave of particle 1.

Complementarily the equation of motion of particle 2 \eref{p2} involves the shock wave field $F^{\m\n}_{1sw}$. This means that this equation makes sense only if it is regarded as a {\it distributional} differential equation. Notice that, since particle 2 never hits the singularity string of particle 1, the term $F^{\m\n}_{1sing}$ drops out also from equation \eref{p2}.
Without facing the problem of its general solution we observe that at each instant $t$ in which particle 2 crosses the shock-wave plane, this equation foresees that the four-momentum $p^\bt_2(t)$ of particle 2 jumps discontinuously by an amount that can be calculated integrating both sides of equation \eref{p2} between $t-\dl$ and $t+\dl$, and sending then $\dl$ to zero. By inspection  the resulting jump of $p_2^\bt(t)$ equals precisely the opposite of \eref{mix}, so that the total four-momentum \eref{mix12} is conserved.

In conclusion a system of two massless charged particles admits well-defined equations of motion - \eref{p1} and \eref{p2} - that are perfectly compatible with local and total energy-momentum conservation. The analysis above extends in a straightforward way to a generic system of massless and massive charges in presence of an external field.

\section{Summary and open problems}\label{sao}

As we have shown, relying on requirements 1)-4) and on the {\it regularity paradigm}, the dynamics of a system of classical massless interacting charged particles, also in presence on an external field, can be formulated in a consistent way.
The electromagnetic field created by the particles - following bounded or unbounded trajectories - is given respectively by \eref{bound} and \eref{unbound}. The particles themselves must obey standard Lorentz-equations, see \eref{dpext}, \eref{p1} and \eref{p2}, as if the self-field could be ignored.  Since these equations are of second order in time derivatives, contrary to the Lorentz-Dirac equation \eref{ldf} their solutions entail no unphysical properties, {\it e.g.}  causality violation in terms of a {\it pre-acceleration}, see for example \cite{Ro}.

The cornerstone of our procedure was the construction of a well-defined energy-momentum tensor, and  - by construction - the total four-momentum is locally conserved. For unbounded trajectories the total four-momentum $P^\bt$ is finite and we gave explicit  expressions, see \eref{ggg} and \eref{mix12}. For bounded trajectories the total four-momentum is generally infinite -  the particle being eternally accelerated the field \eref{bound} generically does fall off at infinity only as $1/r$   -  but local four-momentum conservation still holds, {\it i.e.} $\pa_\a(\mathbb T^{\a\bt}_{em}+T_p^{\a\bt})=0$.

According to our construction the electrodynamics of massless charges is uniquely determined, once we accept the requirements 1)-4) and the {\it regularity paradigm}.
Since the former have a robust physical motivation, a physically inequivalent formulation of this dynamics must renounce to the latter. It is in this less stringent framework that the self-force $f_0^\m$ in \eref{f0}, derived independently also in \cite{KS} and diverging as $w^2\ra 0$, might regain an independent life. According to this framework the particle is subject to the Lorentz-equation \eref{post}, that we write as (for simplicity we ignore the external field, as its inclusion is straightforward, see Section \ref{efa})
\be\label{ks}
\frac{dp^\m}{d\la}=\frac{e^2 (-w^2)^{1/4}}{4\pi}\, f^\m_0.
\ee
To check if this equation is compatible with four-momentum conservation we still resort to the requirements 1)-4). Since we have already constructed an energy-momentum tensor $T^{\a\bt}_{em}$ \eref{temh} satisfying these requirements - in particular $\pa_\a T^{\a\bt}_{em}=0$ - all possible modifications $T^{\a\bt}_{em}+D^{\a\bt}$ of this tensor are characterized by finite local counterterms $D^{\a\bt}$ of the form \eref{dd} - supported thus on the singularity surface - subject to the condition \eref{dcal}
\be\label{dcal1}
 \pa_\a{ D}^{\a\bt}  =\int \!{\cal G}^\bt\, \dl^4(x-y(\la))\,d\la.
\ee
To reproduce equation \eref{ks} in compatibility with local four-momentum conservation (see \eref{dtloc}-\eref{eqm}) the vector ${\cal G}^\bt$  must be given by
\be\label{gmf0}
{\cal G}^\bt=-\frac{e^2 (-w^2)^{1/4}}{4\pi}\, f^\bt_0.
\ee
The problem is therefore reduced to the existence of a tensor $D^{\a\bt}$ of the form \eref{dd}, whose four-divergence satisfies \eref{dcal1} with ${\cal G}^\bt$ given by \eref{gmf0}. A possible candidate, with all correct invariance properties, is {\it e.g.}
\be\label{dks}
D^{\a\bt}=  \int\!\left(u^\a {\cal G}^\bt+ u^\bt {\cal G}^\a\right)\dl^4(x-\G(\la,b))\,dbd\la.
\ee
Using that when acting on  $\dl^4(x-\G(\la,b))$ one has the identification
$u^\a \pa_\a=-\pa/\pa b$, one obtains
\be
\pa_\a D^{\a\bt}= \int\! {\cal G}^\bt \dl^4(x-y(\la))\,d\la+
 \int\! u^\bt {\cal G}^\a \pa_\a\dl^4(x-\G(\la,b))\,dbd\la.
\ee
The first term is actually precisely of the desired form \eref{dcal1}, while the second term is not supported on the world-line, but rather on the singularity surface, and hence the tensor \eref{dks} is not the one we search for. It is clear that, once one renounces to the regularity paradigm, \eref{dks} is far from being the unique permitted candidate, but the example worked out above illustrates that it seems rather difficult - we think impossible - to find a $D^{\a\bt}$ that is supported on the singularity surface and satisfies \eref{dcal1}. The existence of such a  $D^{\a\bt}$, if any, should indeed follow from a - not yet discovered - magic hidden property of the self-force $f_0^\m$ \eref{f0}.
On the other hand the addition ``by hand'' of a complicated and singular counterterm $D^{\a\bt}$ - that apparently has nothing to do with the original energy-momentum tensor \eref{t0}, or its regularized version \eref{the} - would appear rather artificial. Obviously a more exhaustive research in this direction is needed to settle definitely the problem.

Our whole treatment relies on the {\it retarded} electromagnetic field \eref{bound} - a prejudice based on causality; in other words we insist on the field propagating from the particle to the space-time point where the field is observed, and not the opposite. This means that, as in Electrodynamics of massive particles based on the standard Li\'enard-Wiechert-field \eref{lw}-\eref{rt}, time reversal invariance is still {\it spontaneously} broken. There is however a fundamental difference, with this respect, between massive and massless particles: in the time-reversed (unphysical) picture, in the first case the field propagates from infinity to the particle and the particle {\it absorbs} radiation, instead of emitting it, while in the second case the field propagates again from infinity to the particle, but the particle does neither absorb nor emit radiation. Correspondingly in presence of an external field in the
time-reversed picture a massless particle follows a physically {\it allowed} trajectory, while a massive one follows a trajectory that is non-physical, because radiation reaction would increase its energy, instead of lowering it. In summary, a massless charged particle violates still time reversal invariance, but in a weaker sense than massive ones.

Regarding the relation of our work with quantum theory we observe that
in general a consistent quantum formulation of a theory gives rise - in an appropriate limit - to a consistent classical version of that theory.  With respect to the ``fluctuating border'' between classical and quantum electrodynamics we observe, for example, that in the case of massive charges quantum field theory carries a peculiar footprint of classical radiation reaction: it has indeed been shown \cite{HM1,HM2} that the {\it position shift} induced by the (classical) Lorentz-Dirac equation \eref{ldf} can be retrieved directly from Quantum Electrodynamics.

In quantum field theory the main problem related to massless particles regards {\it infrared} divergences: {\it soft} divergences due to massless photons and {\it collinear} divergences due to massless charged particles. In the seminal paper \cite{MS}, based on non-perturbative arguments, it has been argued that - due to these divergences - in four space-time dimensions {\it unconfined} massless charges  can not exist at all. If this were the case, there is no classical limit of a quantum theory through which one could derive the classical dynamics of such particles.

On the other hand, from a perturbative point of view it seems that a consistent quantum theory can be formulated, even if at the moment some fundamental questions - in particular regarding the convergence properties of the Bloch-Nordsieck-Kinoshita-Lee-Naunenberg cancellation mechanism of infrared divergences \cite{BN}-\cite{LN} - are still open \cite{LMc}. As it stands, according to this mechanism the collinear {\it virtual} divergences due to massless charges are canceled by Feynman diagrams corresponding to {\it real} photons emitted/absorbed by the massless charges themselves: this means that in a quantum mechanical perturbative framework massless particles {\it do} emit radiation and that without this radiation  quantum theory could never be consistent. In particular for the cancelation of all collinear infrared divergences {\it both} emission and absorption processes are essential  \cite{LMc}, but clearly only in {\it exceptional} situations the net effect of these processes - from an energetic point of view - is zero. Only in those cases it would be possible to reconcile the occurrence of this quantum-radiation with the {\it absence} of classical radiation, as predicted by our construction.

In conclusion, the perturbative quantum picture seems hardly consistent with our classical construction from basic principles. We are thus led to conclude that, as indicated in the non-perturbative framework of \cite{MS},  unconfined massless charged particles in four space-time dimensionis may exist only at the classical level. As - in the same fashion - massive charged particles in three space-time dimensions appear likewise  to be confined \cite{MS}, the consistency of those particles at the classical level represents an interesting open problem, that we plan to attack in the future.

\paragraph{Acknowledgments.}

This work is supported in part by the INFN Iniziativa Specifica STEFI
and by the Padova University Project CPDA119349.

\vskip0.5truecm

\appendix

\section*{Appendices}

\section{Putative  self-force of a massless particle}\label{sef}

The derivation of \eref{sf} involves a series of successive expansions, that are
more easier to handle if one parameterizes the world-line with the invariant ``proper'' time $\s$ defined in \eref{ds}. To determine $F^{\m\n}_\ve(y)\equiv F^{\m\n}_\ve(y(\s))$ one must first of all determine the retarded time $\s_\ve$ associated - according to \eref{rte} - to the point $x^\m=y^\m(\s)$, {\it i.e.}
\be\label{rete}
(y(\s)-y(\s_\ve))^2=\ve^2.
\ee
As intermediate step it is convenient to introduce a parameter $\D\equiv \D(\s,\ve)$ setting
\be\label{de}
\s_\ve=\s+\D,\quad \quad \D<0,
\ee
where $\D\ra 0$ as $\ve\ra0$. Evaluating the field \eref{freg} at $x=y(\s)$ and multiplying it with $U_\n=dy_\n/d\s$, one can rewrite the regularized self-force \eref{sf} as
\be\label{fme1}
f^\m_\ve= e F^{\m\n}_\ve(y(\s))U_\n=
\frac{e^2}{4\pi(U_\ve L_\ve)}\frac{d}{ d\D}\left(\frac{K^\m}{(U_\ve L_\ve)}\right),
\ee
where we have set
\be
U_\ve^\m=U^\m(\s_\ve),\quad\quad L^\m_\ve=y^\m(\s)-y^\m(\s_\ve),\quad\quad
K^\m=(UU_\ve)L^\m_\ve - (UL_\ve) U^\m_\ve.
\ee
In expanding the quantities appearing in \eref{fme1} in (inverse) powers of $\D$ one can take advantage from the fact that the
quantities $y^{MN}$ defined in \eref{ymn} satisfy the relations (apart from the obvious ones $y^{11}= y^{12}= 0$)
\begin{align}
&y^{13}=1, \quad\quad y^{23}=y^{14}=0,\quad\quad y^{33}=y^{15}=-y^{24},\quad\quad
3y^{16}=15y^{34}=-5y^{25},\nn\\
&y^{17}=8y^{44}+9y^{35},\quad\quad y^{26}=-3y^{44}-4y^{35}.\nn
\end{align}
These relations are derived taking successive derivatives of the identity
$y^{22}=-1$, implied by \eref{ds}.

The main expansions needed are
\begin{align}
-K^\m=&\frac{\D^3}{3}\,y^{1\m} +\frac{\D^4}{12}\,y^{2\m}+\frac{\D^5}{30}\,y^{15}y^{1\m}
+\frac{\D^6}{24}\left(\frac16 \, y^{16}y^{1\m}+\frac{3}{10}\,y^{15}y^{2\m}-\frac{y^{4\m}}{6}\right)\nn\\
&+\frac{\D^7}{120}\left(\frac17\,y^{17}y^{1\m}+\frac13\,y^{16}y^{2\m}+
\frac13\,y^{15}y^{3\m}-\frac13\,y^{5\m}\right)+o\big(\D^8\big),\\
-(U_\ve L_\ve)=&\frac{\D^3}{3!}+\frac{\D^5}{5!}\,y^{15}+\frac{7\D^6}{6!}\,y^{34}
+\frac{\D^7}{7!}\,(15y^{44}+16y^{35})+o\big(\D^8\big).\label{un}
\end{align}
Inserting them into \eref{fme1} one arrives at
\begin{align}
f^\m_\ve=&\frac{e^2}{4\pi}\left(
-\frac{3}{\D^3}\,y^{2\m}-\frac{6}{5\D^2}\,y^{15}y^{1\mu}-
\frac{3}{4\D}\left(\frac{11}{5}\,y^{34}y^{1\m}+y^{15}y^{2\m}-y^{4\m}\right)\right.\nn\\
&\left.+\left(\frac{9}{50}\,\big(y^{15}\big)^2-\frac{18}{35}\,y^{44}
-\frac{22}{35}\,y^{35}\right)y^{1\m} -\frac98\,y^{34}y^{2\m}-\frac25\,y^{15}y^{3\m}+\frac25\,y^{5\m} +o\big(\D\big)\right).\label{fme2}
\end{align}
The final step consists in inserting \eref{de} in \eref{rete} and deriving the expansion of $\D$ in terms of $\ve$ (recall that $\D<0$ and $\ve>0$)
\be
\D=-12^{1/4}\ve^{1/2}+\frac{12^{3/4}y^{15}}{120}\,\ve^{3/2}-\frac{y^{34}}{10}\,\ve^2+
o\big(\ve^{5/2}\big).
\ee
Inserting this expansion in \eref{fme2} one obtains \eref{sf}.

\section{Divergent counterterms}\label{dot}

The derivation of \eref{aab} and \eref{bab} requires to apply the tensors $A^{\a\bt}$ and $B^{\a\bt}$ to a test function $\vp(x)$ and to analyze their behavior  as $\ve\ra 0$. We present the details for $A^{\a\bt}$, indicating for $B^{\a\bt}$  only the main steps.

From \eref{ab}, remembering the definition \eref{lm} and using \eref{rte} to switch from the variable $x^0$ to an independent variable $\la$, after a shift of variables we obtain
\[
A^{\a\bt}(\vp)=-\int\frac {w^2 X^\a X^\bt}{X^0(uX)^3}\,\,\vp(X^0+y^0,\vec x+\vec y)\,d^3x d\la.
\]
In the integral the variables $y^\m$, $u^\m$ and $w^\m$ are evaluated at $\la$ and we have set $X^0=\sqrt{r^2+\ve^2}$, with $r=|\vec x|$, and $\vec X=\vec x$. Thanks to (manifest) reparameterization invariance we can now parameterize the world-line with time, $\la=y^0(\la)\equiv t$, so that
\be\label{time}
u^\m=(1,\vec v),\quad\quad w^\m=(0,\vec a),\quad\quad  w^2=-a^2,
\ee
where we denote the ordinary velocity and acceleration respectively by
$\vec v=d\vec y/dt$ and $\vec a=d\vec v/dt$.
We obtain thus
\be\label{a1}
A^{\a\bt}(\vp)=\int\!\frac {a^2 X^\a X^\bt}{X^0(X^0-\vec v\cdot\vec x)^3}\,\,\vp(X^0+t,\vec x+\vec y)\,d^3x dt.
\ee
The variables $\vec y$, $\vec v$ and $\vec a$ are now evaluated at $t$. As $\ve\ra 0$  the denominator $(X^0-\vec v\cdot\vec x)$ vanishes along the half-line $\vec{x}=b\vec{v}$,  $b>0$, that is the image  at fixed $t$ of the singularity string  $\eref{ga}$; along this line the integral \eref{a1} becomes thus divergent as $\ve\ra 0$. To isolate these divergences it is convenient to  change coordinates from $\vec x \rightarrow (b,q_a)$, $a=1,2$, according to
\begin{equation}
\vec{x}=b\vec{v}+q_{a}\vec{N}_{a},\label{bqVar}
\end{equation}
where $\{\vec{v},\vec{N}_a\}$ is an orthonormal basis  at fixed time, {\it i.e.}
\be\label{bq}
\vec{N}_{a}\cdot\vec{N}_{b}=\delta_{ab},\quad \vec v\cdot\vec v=1,\quad   \vec{N}_{a}\cdot\vec{v}=0, \quad N^i_aN^j_a+v^iv^j=\delta^{ij}.
\ee
In these coordinates the location of the singularity line is simply $q_a=0$ and \eref{a1} becomes indeed
\be\label{a2}
A^{\a\bt}(\vp)=\int\!\frac {a^2 (X^0+b)^3 X^\a X^\bt/X^0}{(q^2+\ve^2)^3}\,\,\vp(t+X^0,\vec y+ b\vec{v}+q_{a}\vec{N}_{a})\,d^2q db dt,
\ee
where now
\be
X^0=\sqrt{q^2+b^2+\ve^2}, \quad\quad \label{xx}
\vec X= b\vec{v}+q_{a}\vec{N}_{a}.
\ee
As $\ve\ra 0$, for $b<0$ the singularities arising at $q=0$ from the denominator in \eref{a2} are compensated by the numerator $(X^0+b)^3$ - the singularity string is indeed a {\it half}-line - so that for what concerns the divergent contributions of $A^{\a\bt}(\vp)$ we can restrict $b$ to {\it positive} values. To perform the explicit expansion of \eref{a2} for $\ve\ra 0$ it is convenient to perform the rescaling $q^a\ra \ve q^a$ and to expand then the numerator of the integrand and the test function in powers of $\ve$. The resulting $q$-integrals become then elementary. The computations are a bit lengthy, but thanks to manifest  Lorentz-invariance of our regularization it is sufficient to perform them for the component $A^{00}$; the tensor $A^{\a\bt}$, being symmetric, can indeed be reconstructed knowing solely $A^{00}$. The result reads $(b>0)$
\begin{align}
 A^{00}(\vp)= -\pi\!&\int\! b^2a^2\! \left\{\!-\frac{4b^2}{\ve^4} +\frac{1}{\ve^2}
\left(2+ 4b \pa^0+b^2 \square\right)\right.\nn\\
&+\left.\ln\ve\!\left(3(\pa^0)^2-\nabla^2+2b \pa^0\square+\frac{b^2}{4}\,\square^2\right)\!\right\}\vp(t+b,\vec y+ b\vec{v})\, db dt+o(1)\nn\\
= \pi\!&\int\! b^2w^2\! \left\{\!-\frac{4b^2}{\ve^4}\,u^0u^0 +\frac{1}{\ve^2}
\left(2+ 4bu^0 \pa^0+b^2u^0u^0 \square\right)\right.\nn\\
&+\left.\ln\ve\!\left(2(\pa^0)^2+\square+2b u^0\pa^0\square+\frac{b^2}{4}\,
u^0u^0\square^2\right)\!\right\}\vp(\G(\la,b)) \,dbd\la +o(1),\label{a00}
\end{align}
where in the second expression we parameterized the world-line again with an arbitrary parameter $\la$ and $\G(\la,b)$ is the singularity surface \eref{Ga}. The expansion \eref{a00} is manifestly reparameterization invariant and, inserting appropriate factors of $\eta^{\a\bt}$ it is straightforward to reconstruct the whole tensor $A^{\a\bt}(\vp)$. Factorizing eventually the test function one recovers \eref{aab}.

Proceeding in the same way for $B^{\a\bt}$, from \eref{ab} one obtains now - instead of \eref{a2}
\begin{align}\nn
B^{\a\bt}(\vp) =\ve^2\!&\int\!\bigg(
\frac{2(X^0+b)^4(wX)u^{(\a}w^{\bt)}}{X^0(q^2+\ve^2)^4}
-\frac{(X^0+b)^3w^{\a}w^{\bt}}{X^0(q^2+\ve^2)^3}\nn\\
&-\left.\frac{(X^0+b)^5(wX)^2u^{\a}u^{\bt}}{X^0(q^2+\ve^2)^5}\right)
\vp(t+X^0,\vec y+ b\vec{v}+q_{a}\vec{N}_{a})\,d^2q db dt,\label{babf}
\end{align}
where $X^\a$ is given in \eref{xx}. Thanks to manifest Lorentz invariance it is again sufficient to expand the component $B^{00}$. Since in \eref{babf} the trajectory is parameterized with time the kinematical quantities are given again by \eref{time}, so that it is only the third term  to give a non-vanishing contribution to this component:
\[
B^{00}(\vp) =-\ve^2\!\int\!\frac{(X^0+b)^5 \big(\vec a\cdot \vec N_c\big)\big(\vec a\cdot \vec N_d\big)\,q^cq^d}{X^0(q^2+\ve^2)^5}\,
\vp(t+X^0,\vec y+ b\vec{v}+q_{a}\vec{N}_{a})\,d^2q db dt.
\]
Due to the pre-factor $\ve^2$ the non-vanishing contributions of $B^{00}(\vp)$ as $\ve\ra 0$  are necessarily supported  on $\G$. Performing the expansion as above one obtains now $(b>0)$
\begin{align}
 B^{00}(\vp)= \pi\!&\int\! b^4\left\{\!-\frac{4}{3\ve^4}\,a^2 -\frac{1}{6\ve^2}
\left(-a^2\square+2a^ia^j\pa_i\pa_j\right)\right.\nn\\
&+\left. \frac{1}{48}\left(-a^2\square+4a^ia^j\pa_i\pa_j\right)\square
\right\}\vp(t+b,\vec y+ b\vec{v})\, db dt+o(\ve)\nn\\
= \pi\!&\int\!b^4 \left\{\!\frac{4w^2}{3\ve^4}\,u^0u^0 -\frac{1}{6\ve^2}
\left(w^2u^0u^0\square +2G^{0\m}G^{0\nu}\pa_\m\pa_\n\right)\right.\nn\\
&+\left.\frac{1}{48} \left(w^2u^0u^0\square +4 G^{0\m}G^{0\nu}\pa_\m\pa_\n\right)\!\right\}\vp(\G(\la,b)) \,dbd\la +o(\ve),\label{b00}
\end{align}
where in the last line we parameterized the world-line again with an arbitrary parameter and the (reparameterization covariant) tensor $G^{\a\bt}$ is given in \eref{gab}. Expression \eref{b00} is easily covariantized, and factorizing the test function one obtains \eref{bab}.

\section{Four-divergence of the regularized energy-momentum tensor}\label{fdo}

Setting
\[
\pa_\a\Theta^{\a\bt}_\ve=\ve^2 \left(\frac{2w^2(wL)}{(uL)^6 } - \frac{(wB)}{(uL)^5}\right)L^\bt \equiv S^\bt,
\]
and proceeding as in Appendix \ref{dot} we get
\[
S^\bt(\vp)=\ve^2\!\int\!\left(\frac {2(X^0+b)^5w^2(wX)}{X^0(q^2+\ve^2)^5}-
\frac{(X^0+b)^4(wB)}{X^0(q^2+\ve^2)^4}\right)\!X^\bt
\vp(t+X^0,\vec y+ b\vec{v}+q_{a}\vec{N}_{a})\,d^2q db dt,
\]
where we used the same notation as in \eref{babf}. Since $S^\bt$ is a vector it is sufficient to expand its time component
\[
S^0(\vp)=\ve^2\!\int\!\left(\frac {2(X^0+b)^5a^2(\vec a\cdot\vec N_c)\,q^c}{(q^2+\ve^2)^5}+
\frac{(X^0+b)^4(\vec a\cdot\dot{\vec a})}{(q^2+\ve^2)^4}\right)
\vp(t+X^0,\vec y+ b\vec{v}+q_{a}\vec{N}_{a})\,d^2q db dt.
\]
Carrying out the computations as in Appendix \ref{dot} we get the expansion (we write it parameterizing the world-line with an arbitrary parameter $\la$)
\[
S^0(\vp)=
\pi\!\int\!\frac{ b^2w^2}{3}\left(\!-\frac{32b}{\ve^4}\,u^0 +\frac{2}{\ve^2}\,
(2bu^0\square +3\pa^0) -\frac12\,(bu^0\square +3\pa^0)\square\right)\vp(\G(\la,b)) \,dbd\la +o(\ve).
\]
From this expression one reads off easily $S^\bt(\vp)$, and factorizing the test function one gets  \eref{d4e}.

\section{Four-momentum integrals}\label{fmi}

To derive \eref{pat} we must integrate the $0\bt$-components of $A^{\a\bt}$ in \eref{ab} over whole space
\[
P^\bt_A(t)=-\int\!\frac{w^2L^0 L^\bt}{(uL)^4}\,d^3x,\quad \quad L^\m(x)=x^\m-y^\m(\la_\ve(x)).
\]
To perform the integral we add two more integrals over the new variables $x^0$ and $\la\equiv\la_\ve(x)$, inserting the $\dl$-functions $\dl(x^0-t)$ and $\dl((x-y(\la))^2-\ve^2)$, see \eref{rte}, and perform then the shift $x^\m\ra x^\m+y^\m(\la)$. In this way we obtain
\begin{align}
P^\bt_A(t)&
=-2\int\!\frac{w^2x^0 x^\bt}{(ux)^3}\,H(x^0)\,\dl\big(x^2-\ve^2\big)\, \dl\big(x^0+y^0(\la)-t\big)d^4xd\la\nn\\
&=
\frac{\pa}{\pa u_\bt}\int\!\frac{w^2x^0}{(ux)^2}\,H(x^0)\,\dl\big(x^2-\ve^2\big)\, \dl\big(x^0+y^0(\la)-t\big)d^4xd\la\nn\\
&=
4\pi\frac{\pa}{\pa u_\bt}\int\!\frac{w^2x^0}
{(u^0x^0)^2-|\vec u|^2\, r^2}\,H(x^0)\,\dl\big(x^2-\ve^2\big)\, \dl\big(x^0+y^0(\la)-t\big)\,dx^0r^2drd\la\nn\\
&=
-\frac{8\pi}{\ve^4}\int\!\frac{w^2x^0\left((x^0)^2,r^2\vec v \right)}
{(u^0)^3}\,H(x^0)\,\dl\big(x^2-\ve^2\big)\, \dl\big(x^0+y^0(\la)-t\big)\,dx^0r^2drd\la,
\label{pfin}
\end{align}
where the variables $w^\m$, $u^0$ and $\vec v=\vec u/u^0$  are evaluated at $\la$.
In the second line we introduced a {\it formal} derivative w.r.t. a generic vector $u^\bt$ - not subject to $u^2=0$ - to get the third line we performed the integral over angles
\[
\int \!\frac{d\Omega}{(ux)^2}= \frac {4\pi}{ (u^0x^0)^2-|\vec u|^2\, r^2},  \quad \quad    r=|\vec x|,
\]
and finally we swapped the derivative $\pa/\pa u_\bt$ with the integral sign and
enforced  eventually the constraint $u^2=0$. Integrating out the $\dl$-functions - using reparameterization invariance to chose $\la=y^0(\la)$ - \eref{pfin} reduces to \eref{pat}.


\begin{thebibliography}{1}
\bibitem{unb} In particular for unbounded motions, in which case \eref{dd} is of compact spatial support, integrating equation \eref{ddab} over whole space one finds that the (finite) total four-momentum $P_D^\bt=\int D^{0\bt}d^3x$, associated to $D^{\a\bt}$, is constant. But since for large negative times the acceleration $w^\m$ has a fast decay, one has actually $P_D^\bt=0$, see the examples \eref{d0}-\eref{d2}.
\bibitem{unb1} Since for unbounded motions the singularity curve $\vec \G$ \eref{ga} is of compact spatial support, for sufficiently large $R$ the tensor  $A^{\a\bt}$ is a $C^\infty$-function, so that the $\ve$-regularization can be removed.
\bibitem{lead} In reference \cite{K} only the presence of a leading divergence resembling \eref{paa} has been revealed, while the subleading divergence and the finite term of \eref{patt} - both being indispensable for total four-momentum conservation and for Lorentz-covariance - have been overlooked.
\end{thebibliography}
\end{document}